\RequirePackage[2020-02-02]{latexrelease}
\documentclass[amscls]{revtex4}
\usepackage{amsmath}
\newcommand{\be}{\begin{equation}}
	\newcommand{\ee}{\end{equation}}
\newcommand{\ba}{\begin{eqnarray}}
	\newcommand{\ea}{\end{eqnarray}}
\begin{document}

\title{Affine Connections in Quantum Gravity and New Scalar Fields}

\author{Kaushik Ghosh\footnote{E-mail ghosh{\_}kaushik06@yahoo.co.in}}
\affiliation{Vivekananda College (University of Calcutta), 
269, Diamond Harbour Road, Kolkata - 700063, India}

\maketitle

\section*{Abstract}

In this manuscript, we will discuss the construction of covariant derivative operator in
quantum gravity. We will find it is more perceptive alternative to use affine connections 
more general than metric compatible connections in quantum gravity. 
We will demonstrate this using the canonical quantization procedure.
This is valid irrespective of the presence and nature of sources. 
The conventional Palatini and metric-affine formalisms, where the actions
are linear in the scalar curvature with metric and affine connections being the 
the independent variables, are not much suitable to 
construct a source-free theory of gravity with general affine connections. 
This is also valid for many minimally coupled interacting theories where sources 
only couple with metric by using  the Levi-Civita connections
exclusively. We will discuss potential formalism of affine connections 
to introduce affine connections
more general than metric compatible connections in gravity.
We will also discuss possible extensions of the actions 
for this purpose. General affine connections introduce new fields
in gravity besides metric.   
In this article, we will consider a simple potential formalism with 
symmetric affine connections and symmetric Ricci tensor.
Corresponding affine connections introduce two massless scalar fields.
One of these fields contributes a stress-tensor 
with opposite sign to the sources of Einstein's equation
when we state the equation using the Levi-Civita connections.
This means we have a massless scalar field with negative stress-tensor in 
Einstein's equation. This field brings us beyond strict local Minkowski geometries.
These scalar fields can be useful to explain inflation and dark energy.

\vspace{0.5cm}

Key-words: Affine connections, quantum cosmology, scalar fields, dark energy, inflation



\vspace{0.5cm}

PACS: 98.80.Qc, 98.80.Cq, 95.36.+x, 04.50.Kd, 02.40.-k,




\newpage

\section*{I. Introduction}

It is now accepted in modern cosmology that the matter and radiation
dominated era of cosmic epoch is sandwiched between two periods of cosmic
acceleration: inflation and dark energy [1,2,3,4,5,6].
The inflationary scenario is based on some scalar field called inflation.
The term dark energy is reserved for an unknown form of energy which 
not only has not been detected directly, but also does not cluster as ordinary matter does.
It is hypothesized to have negative pressure to explain present
cosmic acceleration.
Besides these, we also need dark matter which does not mediate electromagnetic
interactions and are observed by their gravitational effects [6,7,8,9]. 
Present data from different sources, such as the Cosmic Microwave Background Radiation 
(CMBR) and supernovae surveys, seem to indicate that the energy 
composition of the universe consists of $20 \%$  dark matter,
$76 \%$ dark energy and the rest ordinary baryonic matter.  
The simplest candidate for dark energy is the cosmological constant
$\Lambda$ with constant energy density although we will have to 
explain its small magnitude [6,10]. There are two approaches to
explain cosmic acceleration as alternatives to the cosmological constant model. 
The first is to supplement the source stress-tensor part of Einstein's equation by
specific forms of stress-tensor with negative pressure. Among various
models, cosmons or quintessence, k-essence and perfect fluid models
are mostly studied [11,12,13]. The second approach to 
explain dark energy is to modify the gravitational part of Einstein's equation.
Examples are the so-called $f(R)$ gravity [9,14], scalar-tensor
theories [15,16] and braneworld models [17,18]. The
modified gravity models are more strongly constrained than the modified
matter models from astrophysical and cosmological observations. 
Despite a lot of efforts during the last two decades, it is fair to say
that we are yet to explain the origin of the inflation, cosmological
constant, dark energy and dark matter. 
The $\Lambda$-Cold Dark Matter ($\Lambda$CDM) model can not explain the
origin of the above set of fields.

There have been fair amount of works to extend the Einstein-Hilbert
action to construct a renormalizable theory of quantum gravity 
[19,20]. It was also found that when quantum corrections 
are taken into account, the effective low energy gravitational action admits higher order curvature invariants 
[21,22,23,24]. The gravitational part of Einstein's equation changes when we use such actions. 
These actions have been useful to construct various modified theories of
gravity mentioned above.

In this manuscript, we will try to find if quantum gravity can introduce additional
fields besides metric. We will discuss this with reference to the construction of 
covariant derivative operator in quantum gravity.
We can introduce a metric in a spacetime manifold provided
it satisfies a few very general topological conditions [25]. 
To construct covariant differential equations for different tensor fields, we have to
introduce additional structures in spacetime. These structures are known as connections.
They can be introduced in a differential manifold independent of metric [26,27].
In gravity, we only consider affine connections which we will describe briefly in 
section:II. In this article we will find it is a perceptive alternative to use affine connections 
more general than metric compatible affine connections in quantum gravity. 
We will demonstrate this using the canonical quantization procedure.
General affine connections introduce additional fields in the theory that can
be useful to explain cosmological observations mentioned before and can 
introduce new effects. These fields are non-localized similar to dark energy and inflation.
In this article, we have considered a simple case with symmetric Ricci tensor.
Corresponding affine connections are symmetric and introduce two massless scalar fields. One of these gives 
a stress tensor with opposite sign to the sources of Einstein's equation 
in the classical theory when we state the equation using the Levi-Civita connections.   
This means we have a source with negative stress tensor in the familiar
Einstein equation. These scalar fields can be possible candidates for
dark energy and inflation.  
General affine connections break strict local Minkowski structure of spacetime.
Global splitting of spacetime into space and time no longer remains exact in such spaces. 
This is an important issue in quantum gravity [28,29,30,31]. Finding the particle interpretations 
and other possible interactions of the above fields are non-trivial problems 
without local Minkowski structure. However, experiments suggest that such 
effects are very small in the present universe and the corresponding scalars mentioned
above are also small. Their effects can be observed in large scale phenomena
like the present cosmic acceleration. This is also consistent with the smallness of the 
parameters like the cosmological constant required
to explain dark energy in some models mentioned at the beginning. 
The effects of these fields will be large in the quantum gravity domain.
We note that vacuum fluctuations of these fields can also be useful in cosmology.
The plan of the paper is then as follows.

We will give a brief description of differential geometry with
general affine connections in section:II. This is required to
construct a theory of gravity with general affine connections.
We will start with a definition of general affine connections [26,27,28,32,33,34].
In the rest of this article, we will deal with only affine connections
and denote general affine connections by affine connections or connection coefficients. 
The well-known Levi-Civita connections are a special set of affine connections
which obey additional conditions.
Affine connections may be asymmetric in the lower indices and need not
to be compatible with metric or any symmetric second rank covariant tensor [26].
Affine connections give an additional third rank tensor besides metric.
In this article, we denote this field by ${C^{\alpha}_{~\mu \nu}}$.
Antisymmetric part of this tensor in the covariant indices 
is half of torsion [26], and give antisymmetric fluctuations
in affine connections away from the Levi-Civita connections.
Symmetric fluctuations of affine connections off the Levi-Civita
connections are described by the symmetric part of ${C^{\alpha}_{~\mu \nu}}$ in the lower
indices and this field is not considered in conventional 
theories of quantum gravity even with sources.
In general, this field can be finite when torsion is so. 
This is evident from Eqs.(25-28). 
The symmetric part of ${C^{\alpha}_{~\mu \nu}}$ in the lower
indices can introduce new symmetric second rank covariant tensors
and scalars in gravity besides metric.

In section:III, we will find it is more appropriate to use affine connections more general than
metric compatible connections in quantum gravity.
We will use the canonical quantization procedure and the 
Arnowitt-Deser-Misner (ADM) formalism [29] to show this.
This is valid irrespective of the presence and nature of sources.
This is a general mathematical issue which will be there, in a theory of 
quantum gravity which is not a quantum field theory in a fixed background, 
provided some components of metric can be taken as independent 
variables in a neighborhood of the spacetime manifold. 
This can be done around any regular point of the spacetime manifold [35,36,37].
We will also use the general metric-metric commutators to illustrate this issue [28].

In section:IV, we will discuss the construction of a suitable action, 
where metric and affine connections are the independent variables, 
for a quantum theory of gravity. 
We will first consider the Palatini and metric-affine formalisms [28,38]. 
In these formalisms, the gravitational Lagrangian density is given by
the scalar curvature obtained from the corresponding general affine connections.
The Palatini action principle leads to metric compatible Levi-Civita connections.
This is not suitable to construct quantum gravity where we need to remove
the metric compatibility condition even in the source-free theory. 
The metric-affine theory without sources can only introduce a limited version of non-metricity 
due to projective invariance. Thus, it is appropriate to extend these formalisms to introduce non-metricity.
We will have to extend both the formalisms
if we want to have dynamics for ${C^{\alpha}_{~\mu \nu}}$.

We will discuss a potential formalism of affine connections in section:V. 
By potential formalism we mean a formalism where ${C^{\alpha}_{~\mu \nu}}$
is derived from a tensor of lower rank. This is case with the Levi-Civita
connections that are derived from metric.
This formalism can introduce finite and dynamic affine connections
in the Palatini and metric-affine gravity even when there is no source
and introduces a new second rank covariant tensor in the theory.
We can also modify the actions by introducing higher order
curvature invariants or use theories like Palatini $f(R)$ gravity and metric-affine $f(R)$ gravity
to introduce dynamical ${C^{\alpha}_{~\mu \nu}}$.
In section:VI we will discuss simple applications of the potential formalism mentioned above. In
section:VII we will discuss extension of the Einstein equation with finite ${C^{\alpha}_{~\mu \nu}}$ and the issue of stability.

\section*{II. Affine Connections and Covariant Derivatives}

In this section, we will briefly discuss differential geometry of of curved spaces
with general affine connections. We will also mention the generalization of the Einstein-Hilbert
action with general affine connections. This formalism is known
as the metric-affine theory of gravity [38].  
Four properties of ordinary derivatives and transformation rules of tensors under the change of coordinate
systems are used to construct covariant derivatives in
a general manifold [26,28]. Consequently, covariant derivatives in a curved spacetime inherit
these properties. We state them in the following:

(I) The linearity property: 
${{\nabla}_{\mu}}[a {p^{.. \alpha_i ..}_{.. \beta_j ..}} + b {q^{.. \alpha_i ..}_{.. \beta_j ..}}] 
= a {{\nabla}_{\mu}}{p^{.. \alpha_i ..}_{.. \beta_j ..}} + 
b {{\nabla}_{\mu}}{q^{.. \alpha_i ..}_{.. \beta_j ..}}$.
Where, $a,b$ are two constants and ${p^{.. \alpha_i ..}_{.. \beta_j ..}},
{q^{.. \alpha_i ..}_{.. \beta_j ..}}$ are two well-behaved tensor fields.

(II) For a well-behaved scalar field $f$, and a vector field $t^{\mu}$, 
$t(f) = {t^{\mu}}{{\nabla}_{\mu}}(f)$. Here, $t(f)$ denotes directional derivative of the 
scalar field in the direction of the vector field.

(III) The Leibnitz rule:

\be
{{\nabla}_{\mu}}[{p^{.. \alpha_i ..}_{.. \beta_j ..}} {q^{.. \alpha_i ..}_{.. \beta_j ..}}] 
= [{{\nabla}_{\mu}}{p^{.. \alpha_i ..}_{.. \beta_j ..}}] {q^{.. \alpha_i ..}_{.. \beta_j ..}} 
+  {p^{.. \alpha_i ..}_{.. \beta_j ..}} [{{\nabla}_{\mu}}{q^{.. \alpha_i ..}_{.. \beta_j ..}}]
\ee

\noindent{Here, ${p^{.. \alpha_i ..}_{.. \beta_j ..}}, {q^{.. \alpha_i ..}_{.. \beta_j ..}}$ are two arbitrary well-behaved tensor fields.}

(IV) Commutativity between contraction and covariant derivative:

\be
{{\nabla}_{\mu}}[C({p^{.. \alpha_i ..}_{.. \beta_j ..}})] = C({{\nabla}_{\mu}}[{p^{.. \alpha_i ..}_{.. \beta_j ..}}])
\ee

\noindent{Here, $C(~)$ denotes contraction operation between upper and lower indices
using the Kronecker's delta. These properties and tensorial character of covariant derivatives lead to the 
following property:}

(V) We define torsion tensor through the following relation:

\be
[{{\nabla}_{\mu}}{{\nabla}_{\nu}} - {{\nabla}_{\nu}}{{\nabla}_{\mu}}]f = 
- {T^{\alpha}_{~\mu \nu}}{\nabla_\alpha}f
\ee

\noindent{where, $f$ is a well-behaved scalar field. ${T^{\alpha}_{\mu \nu}}$ is torsion
tensor.}

We can construct a covariant derivative operator having properties (I - IV):

\be
{{\nabla}_{\mu}}{A_{\nu}} = {{\ddot{\nabla}}_{\mu}}{A_{\nu}} -  
{{\Theta}^{\alpha}_{~\mu \nu}}{A_{\alpha}}
\ee

\noindent{Here, ${{\Theta}^{\alpha}_{~\mu \nu}}$ are connections. 
In general relativity, we choose ${{\ddot{\nabla}}_{\mu}} \equiv {\partial_{\mu}}$, where 
${\partial_{\mu}}$ is ordinary partial derivative
and we have: $t(f) = {t^{\mu}}{\partial_{\mu}}(f)$. Corresponding connections
for which ${{\nabla}_{\mu}}$ is a covariant derivative operator
and satisfy the above conditions are known as affine connections or connection coefficients [26,28]. 
In this article, we have used the terms affine connections or connection coefficients
to denote the most general affine connections. The well known
Levi-Civita connections are a special set of affine connections that obey
additional conditions to be mentioned below. 
Tensorial character of covariant derivatives impose the following transformation
rule on affine connections:}

\be
{{\bar{\Theta}}^{\alpha}_{~\mu \nu}} =
{{\partial{\bar{x}^{\alpha}}} \over {\partial{x^{\lambda}}}}
{{\partial{x^{\kappa}}} \over {\partial{\bar{x}^{\mu}}}} 
{{\partial{x^{\tau}}} \over {\partial{\bar{x}^{\nu}}}}{{\Theta}^{\lambda}_{~\kappa \tau}}
+
{{\partial{\bar{x}^{\alpha}}} \over {\partial{x^{\lambda}}}}
{{\partial^{2}{x^{\lambda}}} \over {\partial{\bar{x}^{\mu}} \partial{\bar{x}^{\nu}}}}
\ee

\noindent{We can make ${{\Theta}^{\alpha}_{~\mu \nu}}$
symmetric in the lower indices. Corresponding connections are  
known as the Chistoffel symbols [28]. We can introduce additional conditions on the Chistoffel
symbols. In general relativity, we introduce the Levi-Civita connections through
the following metric compatibility conditions [28]:}

\be
{\nabla_{\mu}}[{g_{\alpha \beta}}] = 0
\ee

\noindent{We have the following expressions for them:}

\be
{{\Theta}^{\alpha}_{~\mu \nu}} = {\Gamma^{\alpha}_{~\mu \nu}} = {1 \over 2}
[{\partial_{\mu}}(g_{\kappa \nu}) + {\partial_{\nu}}(g_{\mu \kappa})
- {\partial_{\kappa}}(g_{\mu \nu})]{g^{\alpha \kappa}}
\ee

\noindent{The above expression shows that the Levi-Civita connections are dependent
on partial derivatives of metric and are symmetric in the lower indices. 
Note that, we can have other solutions of Eq.(6) by introducing additional fields
besides metric. We will use such connections below. 
Torsion is zero when connections are symmetric in the lower
indices. Thus, the right hand side of Eq.(3)
is zero when we use the Levi-Civita connections. 
The familiar solutions of Einstein's equation in general relativity satisfy
the torsion-free condition [28]. In this article,
we will show that we have to use affine connections more general than metric compatible
connections in quantum gravity.}

We next consider the metric-affine theory of gravity. 
The source-free action is given by [28,38]:

\be
S = {\int}{\sqrt{-g}}{R}{\bf e} 
\ee

\noindent{Where, $g$ is the determinant of metric,
${\sqrt{-g}}{\bf e}$ is the natural volume element associated with
metric and  $R$ is the scalar curvature. In this formalism,
both metric and affine connections are the independent variables. 
We have the following expression for covariant derivative operator [28]:}

\be
{\nabla_{\mu}}{A_{\nu}} = {{\nabla'}_{\mu}}{A_{\nu}} -  
{C^{\alpha}_{~\mu \nu}}{A_{\alpha}}
\ee

\noindent{Here, ${C^{\alpha}_{~\mu \nu}}$ is an arbitrary 
well-behaved field. It can be symmetric or asymmetric in the lower indices. 
${{\nabla'}_{\mu}}$ is a given 
covariant derivative which have properties (I - IV) and obey the torsion-free
condition. We choose ${{\nabla'}_{\mu}}$ to be given by the following expression:}

\be
{{\nabla'}_{\mu}}{A_{\nu}} = {\partial_{\mu}}{A_{\nu}} -  
{\Gamma^{\alpha}_{~\mu \nu}}{A_{\alpha}}
\ee

\noindent{Here, ${\Gamma^{\alpha}_{~\mu \nu}}$ are the Levi-Civita connections 
associated with metric $g_{\mu \nu}$. With this choice of ${{\nabla'}_{\mu}}$, 
${C^{\alpha}_{~\mu \nu}}$ is a third rank tensor. This follows from the transformation
properties of affine connections given by Eq.(5) and the definition of 
Levi-Civita connections given by Eq.(7). Note that, ${{\nabla'}_{\mu}}$
commutes with index raising/lowering operations although ${{\nabla}_{\mu}}$ does not.
This is another advantage of using Eqs.(9,10).}

A restricted version of the metric-affine action mentioned above
is obtained if we only consider ${C^{\alpha}_{~\mu \nu}}$ that are
symmetric in the lower indices. This is known as the 
Palatini formalism [28]. In many cases, the symmetric part of 
affine connections in Eq.(9) can be given by the following expression [26]:

\ba
{S^{\alpha}_{~\mu \nu}} & = & {\Gamma^{\alpha}_{~\mu \nu}} + {{\tilde{C}}^{\alpha}_{~\mu \nu}} \\ \nonumber
& = & {1 \over 2}
[{\partial_{\mu}}(b_{\kappa \nu}) + {\partial_{\nu}}(b_{\mu \kappa})
- {\partial_{\kappa}}(b_{\mu \nu})]{\tilde{b}^{\alpha \kappa}}
\ea

\noindent{Where, ${\Gamma^{\alpha}_{~\mu \nu}}$ are the Levi-Civita connections and given
by Eq.(7). ${{\tilde{C}}^{\alpha}_{~\mu \nu}}$ is the symmetric part of 
$C^{\alpha}_{~\mu \nu}$ in the lower indices. 
$b_{\mu \nu}$ is a non-singular symmetric covariant tensor and can be
expressed as:}

\be
b_{\mu \nu} = g_{\mu \nu} + a_{\mu \nu}
\ee

\noindent{The inverse of $b_{\mu \nu}$, ${\tilde{b}^{\mu \nu}}$,
is a contravariant tensor [26] and can be
expressed as ${\tilde{b}^{\mu \nu}} = g^{\mu \nu} + d^{\mu \nu}$. 
Note that ${\tilde{b}^{\mu \nu}}$ is different from ${b}^{\mu \nu}$
and $S^{\alpha}_{~\mu \nu}$ satisfy the compatibility conditions: $b_{\mu \nu | \alpha} = 0$,
where the bar denotes covariant derivative with connections $S^{\alpha}_{~\mu \nu}$. 
We will later find that  the Ricci tensor is symmetric when affine connections 
are exclusively given by Eq.(11). However, Eq.(11) is not valid when the Ricci tensor
is not symmetric. The antisymmetric part of ${C^{\alpha}_{~\mu \nu}}$ 
in the lower indices gives half of torsion tensor [26].
In the metric-affine and Palatini formalisms, metric and respective 
${C^{\alpha}_{~\mu \nu}}$ are taken to be the independent
variables. The Riemann curvature tensor is now defined by the following expressions, 
[26,28,39]:}

\ba
({{\nabla_{\mu}}{\nabla_{\nu}}} - {{\nabla_{\nu}}{\nabla_{\mu}}}){A^{\alpha}_{~\beta}}
& = & - {R_{\mu \nu \kappa}^{~~~~\alpha}}{A^{\kappa}_{~\beta}} +
{R_{\mu \nu \beta}^{~~~~\kappa}}{A^{\alpha}_{~\kappa}} 
- {T^{\kappa}_{~\mu \nu}}{\nabla_{\kappa}}{A^{\alpha}_{~\beta}} \\ \nonumber
R_{\mu \nu \alpha}^{~~~~\kappa} & = & {{R'}_{\mu \nu \alpha}^{~~~~\kappa}} + 
2{{\nabla'}_{[\nu}}{C^{\kappa}_{~\mu] \alpha}}
+ 2 [{C^{\lambda}_{~[\mu |\alpha|}} {C^{\kappa}_{~\nu |\lambda|]}}]\\ \nonumber
{T^{\kappa}_{~\mu \nu}} & = & {C^{\kappa}_{~\mu \nu}} - {C^{\kappa}_{~\nu \mu}}
\ea

\noindent{Here, ${R'}_{\mu \nu \alpha}^{~~~~\kappa}$ is the Riemann curvature tensor 
associated with the derivative ${\nabla'_{\mu}}$ in Eq.(10), and is given by the familiar
expression in terms of ordinary partial derivatives of ${\Gamma^{\alpha}_{~\mu \nu}}$ [28].
${T^{\kappa}_{~\mu \nu}}$ is torsion tensor. The second equation is always valid when
${\Gamma^{\alpha}_{~\mu \nu}}$ is symmetric in the lower indices.
The curvature scalar is obtained by usual contractions and is given by Eq.(22).}

\section*{III. Affine Connections in Quantum Gravity}

We now consider quantization of gravity by using the canonical quantization procedure.
Canonical quantization is important to find the particle spectrum when we
quantize a classical theory. In the canonical quantization of gravity,  
metric becomes operator on a Hilbert space. We represent such operators by carets.
Affine connections present in the covariant derivatives act on the tensor 
operators and we represent them also by the symbols: 
${\hat{\Theta}^{\alpha}_{~\mu \nu}}$. Affine connections will contain components of 
metric and their spacetime derivatives and also other fields as evident from the previous discussions.

In a Hamiltonian formulation, induced metric on a
set of constant time surfaces is used as dynamical variable. 
The induced metric on a set of constant time surfaces is given by:

\be
h_{\mu \nu} = g_{\mu \nu} + {n_{\mu}}{n_{\nu}} 
\ee

\noindent{where $n_{\mu}$ is the
unit normal to the constant time surfaces. An expression for conjugate
momenta is given by Eq.(18). We presently use the symbols $\hat{h},\hat{\pi}$ 
to denote the corresponding collection of canonical operators.
In general, the Levi-Civita connections contain metric and time derivative of metric components
and hence, will depend on the canonical conjugate variables $(\hat{h},\hat{\pi})$.
We now express covariant derivative operator in the following form:}

\be
{{\hat{\nabla}'}_{\mu}}{\hat{A}_{\nu}} 
= [{{\partial}_{\mu}} -  
{{\hat{\Gamma}}^{\alpha}_{~\mu \nu}}(\partial{\hat{g}}, \hat{g})]{\hat{A}_{\alpha}} \\ \nonumber
= [{{\partial}_{\mu}} -  
{{\hat{\Gamma}}^{\alpha}_{~\mu \nu}}(\hat{\pi}, \hat{h})]{\hat{A}_{\alpha}}
\ee

\noindent{Here, ${\hat{\Gamma}^{\alpha}_{~\mu \nu}}$ are operator version of the Levi-Civita
connections. We adopt the following operator ordering in connection coefficients. Whenever
there appears a product between partial derivatives of metric and metric itself, the 
partial derivative is kept as the first term and metric is kept as the second term.
The ordering of the operators $(\hat{h},\hat{\pi})$ 
in ${\hat{\Gamma}^{\alpha}_{~\mu \nu}}$
is given to be the same as that written in the above equation, \textit{i.e}, 
$\hat{h}$ is kept as the successor of $\hat{\pi}$.}

We next consider the operator: ${q^{\mu}}{{\hat{\nabla}'}_{\mu}}{q^{\nu}}$,
where $q^{\mu}$ is a vector field acting as $q^{\mu}{\hat{I}}$ on the Hilbert space.
This operator contains canonical conjugate pairs of variables when we choose affine connections
to be given by the Levi-Civita connections. In this case, 
we will have the following expression:

\be
[{q^{\mu}}{{\hat{\nabla}'}_{\mu}}{q^{\nu}}]\left|\Psi\right\rangle
\neq 0
\ee

\noindent{remaining valid in a given state $\left|\Psi\right\rangle$ with an arbitrary
well-behaved vector field ${q^{\mu}}$. We will not
have a complete set of states for which the expectation value of the operator in the \textit{l.h.s}
is zero with negligible fluctuations for all well-behaved vector fields.
This will be valid only in the classical limit,
and is a subject similar to the familiar Ehrenfest's theorems in non-relativistic quantum mechanics.
Similar discussions will remain valid even if we choose 
affine connections to be given by the operator versions of Eqs.(9,10). 
In general, affine connections will contain canonical pairs of variables from metric sector to have proper
classical limit of the Levi-Civita connections, and the concept of
geodesics will not remain exact for all vector fields 
in a quantum state. This will also remain valid for parallel transport and the notion of 
parallel transport is not exact in a quantum theory of gravity.  
This is expected and indicates that we can use affine connections more
general than the metric compatible connections even in free quantum gravity.}

We now consider the metric compatibility conditions. The metric compatibility conditions
given by Eq.(6) are to be replaced by the operator identity: 
${\hat{\nabla}'}_{\mu}[\hat{g}_{\alpha \beta}] \equiv 0$. The action of 
${\hat{\nabla}'}_{\mu}[\hat{g}_{\alpha \beta}]$ on any
state is zero if connection operators are given by the Levi-Civita connection operators
and we choose the operator ordering same as that mentioned
below Eq.(15). Here, we always keep metric operators as the successors
of the partial derivatives of themselves. Thus, the ordering of the different operators in the quantum
version of the Levi-Civita connections will be the same as that given in Eq.(7).
The same will also remain valid for ${\hat{\nabla}'}_{\mu}[\hat{g}_{\alpha \beta}]$.
Here, $\hat{g}_{\alpha \beta}$ will be kept at the right of the Levi-Civita connections.
We also define the contravariant components of metric as: 
${\hat{g}^{\alpha \kappa}}{\hat{g}_{\kappa \beta}} = {{\delta^{\alpha}_{\beta}}}$. 
This ordering leads to the operator identity: ${\hat{\nabla}'}_{\mu}[\hat{g}_{\alpha \beta}] \equiv 0$
irrespective of the ordering of $(\hat{h},\hat{\pi})$ chosen in the partial derivatives of 
metric components. However, the operator version of metric compatibility
conditions need not be consistent with a canonical quantization condition.
We will demonstrate this in the following.

As mentioned above, in a Hamiltonian formulation we use induced metric on a set of
constant time surfaces as dynamical variable. Thus, $g_{\mu \nu}$ is replaced by:
$h_{\mu \nu} = g_{\mu \nu} + {n_{\mu}}{n_{\nu}}$. Here, $n_{\mu}$ 
is given by $(-N,0,0,0)$, $N$ being the lapse 
function. The contravariant $n^{\mu}$ is given by: ${1 \over {N}}(1,-N^{1},-N^{2},-N^{3})$;
where $N^{i}$ are the shift functions. We have: $g_{00} = {N_k}{N^k} - N^{2}$,
$g_{0i} = N_{i}$ and $N_{i} = g_{ij}{N^{j}}$, [29,37].
The induced metric on the constant time surfaces coincide with the spatial part of $g_{\mu \nu}$
which are expressed as $g_{i j}$. These fields are taken to dynamical variables in general relativity   
and we have the following Poisson brackets:

\be
\left\{{g}_{i j}(t,\vec{x}), {\pi}^{k l}(t,\vec{y})\right\} =
{\delta^{k}_{(i}}{\delta^{l}_{j)}}[{\delta}(\vec{x},\vec{y})]
\ee

\noindent{where, $\vec{x}$ refers to the spatial coordinates and the Poisson
bracket is evaluated at equal time. The delta function is defined without recourse to metric.
The conjugate momentum is a spatial tensor density and is given by:}

\be
{\pi}^{p l} = - {\sqrt{|g_{i j}|}}{({K^{p l}} - K{g^{p l}})}
\ee

\noindent{where $|g_{i j}|$ is the determinant of the spatial metric, 
${K_{p l}} = - {{\nabla'}_{p}}{n_{l}}$ is
the extrinsic curvature of the spatial sections, ${g^{p l}}$ is the
inverse of ${g_{p l}}$ and $K$ is the trace of the extrinsic curvature
taken \textit{w.r.t} ${g_{i j}}$ [28,37]. 
We can also define the Poisson brackets for the lapse and shift functions although
their conjugate momenta vanish giving primary constraints [37].
Thus, in a Hamiltonian formulation with the Einstein-Hilbert action, we have
constraints and we can not naively replace the Poisson brackets by commutators when we try to quantize
the theory [37]. There are two principal approaches to quantize the theory [36,37]. 
In the first approach, 
gauge fixing conditions are introduced to render the complete set of constraints
second class [37]. These conditions also determine the lapse and shift functions. 
We then pick two components of $g_{ij}$ as independent variables 
and quantize these components using standard commutation
relations. We can solve the constraints to evaluate other commutators. The second approach is similar
to the Gupta-Bleuler method used to quantize electrodynamics and was initiated by Dirac [37,40].
In this approach the classical variables are treated as independent variables and the 
constraints are imposed on the quantum states. In this case, we can replace the 
classical Poisson brackets by commutators when we quantize the theory.}

We now demonstrate that it is appropriate to extend the Levi-Civita connections 
and metric compatibility conditions as long as we can regard components of 
spatial metric on the constant time surfaces as independent physical variables 
subjected to usual canonical quantization conditions. We will also find that 
we can not have a Hilbert space on which we can impose the metric compatibility 
conditions when such quantization conditions remain valid.
In the following, we will restrict our attention 
to a neighborhood around a regular point $'x'$. We can extend the neighborhood to
the complete spacetime manifold leaving away singularities and other possible irregular points
associated with the constraints [36,37]. We pick a component of spatial metric, say
$g_{p l}$, as an independent physical variable. We then have the following equal time commutator:

\be
[\hat{g}_{p l}(t,\vec{x}), \hat{\pi}^{p l}(t,\vec{y})] = 
{i}{\delta^{p}_{(p}}{\delta^{l}_{l)}}[{\delta}(\vec{x},\vec{y})]
\ee

\noindent{where, the point $'y'$ belongs to the above mentioned neighborhood
of $'x'$. There will be another such commutator for the other independent
variable. The \textit{r.h.s} of the commutator is taken to be a
distribution that is a spatial tensor density in the spatial coordinates of 
$\hat{\pi}^{p l}(t,\vec{y})$. The \textit{r.h.s} will be replaced by different
expressions when we replace $\hat{g}_{p l}(t,\vec{x})$ by any dependent
component of metric including $g_{0 \mu}$. These terms are determined by 
the secondary constraints, gauge fixing conditions, definitions
of the lapse and shift functions given before and the fundamental commutators 
given by the above equation. Covariant derivatives give changes in a tensor when
we move from one point to a neighbouring point. If the Levi-Civita connections are consistent with
the commutators obtained above, corresponding spatial covariant derivatives
of both sides of any of the commutators \textit{w.r.t} the arguments of metric 
will agree since both sides are equal for all components of metric. 
We now consider Eq.(19). The action of ${\hat{\nabla}'}_{x k}$ on the \textit{r.h.s} is 
same as that on a second rank covariant tensor and will contain spatial partial derivatives 
of the delta function. It will also contain additional terms 
dependent on connections and metric that can also explicitly depend on time 
due to explicit time dependence of the gauge fixing conditions [37]. 
The later terms are of the form: 
$-i[{{\hat{\Gamma}}^{\mu}_{~k p}(t,\vec{x})}{{\hat{X}}^{p l}_{\mu l}}
+ {{\hat{\Gamma}}^{\mu}_{~k l}(t,\vec{x})}{{\hat{X}}^{p l}_{p \mu}}]$, where
$i{{\hat{X}}^{p l}_{\alpha \beta}}$ is the \textit{r.h.s} of Eq.(19) when ${\hat{g}_{p l}(x)}$ is
replaced by general ${\hat{g}_{\alpha \beta}(x)}$.
This covariant derivative is not vanishing in general for all values of $\vec{x}$. 
The left hand side vanishes as can be found from the following expression:}

\be
{{\hat{\nabla}'}_{x k}}\left\{{\hat{g}_{p l}(t,\vec{x})}\right\}{\hat{\pi}^{p l}(t,\vec{y})}
- {\hat{\pi}^{p l}(t,\vec{y})}{{\hat{\nabla}'}_{x k}}\left\{{\hat{g}_{p l}(t,\vec{x})}\right\} = 0
\ee

\noindent{This follows since we are imposing the operator versions of the 
metric compatibility conditions. This can also be seen by 
applying the \textit{l.h.s} of the above equation to any state, introducing
a sum over a complete set of states between the products of the operators
and using the fact that the action of ${\hat{\nabla}'}_{\mu}[\hat{g}_{\alpha \beta}]$ on any
state is zero if connection operators are given by the Levi-Civita connection operators
with operator ordering chosen below Eq.(16). 
Thus, the Levi-Civita connections are not consistent with the canonical commutators
given by Eq.(19). We will consider other operator ordering in 
the Levi-Civita symbols in Appendix:A and we will find that similar inconsistency 
arise in these cases also. Similar situation will remain valid for
any point in the manifold where we can introduce constant time surfaces
and assume the existence of a metric component as an independent field in a neighborhood 
around that point. The above inconsistency will also arise with a different choice
of constant time surfaces. Thus, there will be a multitude of coordinate systems where
we can not use the Levi-Civita connections as connection coefficients
if we impose the quantization condition given by Eq.(19). 
This also indicates that we can not use the Levi-Civita connections as connection coefficients
in all coordinate systems that are diffeomorphic to these coordinate systems
due to the tensorial character of ${C^{\alpha}_{~\mu \nu}}$.}

In Dirac's approach to quantize gravity, all the classical variables are independent 
and we quantize them accordingly.
We can find out $[{\hat{g}_{0 \beta}(t,\vec{x})}, {\hat{\pi}^{p l}(t,\vec{y})}]$
from the definitions of the lapse and shift functions and 
$[{{\hat{N}^{\lambda}}(t,\vec{x})}, {\hat{\pi}^{p l}(t,\vec{y})}] = 0$, 
where $N^{0} = N$, [37]. All spatial components $g_{i j}$ satisfy canonical commutation
relations given by Eq.(19). We will again have the inconsistency
mentioned above when we use the Levi-Civita connections. In this case,
the action of ${\hat{\nabla}'}_{x k}$ on the \textit{r.h.s} of Eq.(19) 
is given by expressions like: 
$i [{{\partial_{x k}}{\delta}(\vec{x},\vec{y})} - 
2 {{\hat{\Gamma}}^{0}_{~k p}}(t,\vec{x}) {{\hat{N}^{p}}(t,\vec{x})}{\delta}(\vec{x},\vec{y}) - 
2 {{\hat{\Gamma}}^{p}_{~k p}}(t,\vec{x}) {\delta}(\vec{x},\vec{y})]$, where we have
taken $p = l$ and there is no sum over the repeated indices.
Also, we can not make the 
Levi-Civita connections consistent with the commutation relation given by Eq.(19)
by introducing additional constraints on the physical Hilbert space. If we demand 
that the action of ${\hat{\nabla}'}_{x k}$ to the \textit{r.h.s} of Eq.(19)
vanishes on the physical Hilbert space, we will have the constraint:

\be
[{{\partial_{x k}}{\delta}(\vec{x},\vec{y})} - 
2 {{\hat{\Gamma}}^{0}_{~k p}}(t,\vec{x}) {{\hat{N}^{p}}(t,\vec{x})}{\delta}(\vec{x},\vec{y}) - 
2 {{\hat{\Gamma}}^{p}_{~k p}}(t,\vec{x}) {\delta}(\vec{x},\vec{y})]|\Psi{\rangle} = 0
\ee

\noindent{Where, we have again taken $p = l$ and there is no sum over the repeated indices. 
There will be other similar constraints associated with other commutators.
These states also satisfy the secondary constraints. We have used the
operator identities ${\hat{\nabla}'}_{\mu}{\hat{g}_{\alpha \beta}} = 0$, 
and the Levi-Civita connections. 
All the above conditions will lead to singular expressions involving
$\delta(\textbf{0})$ and the partial derivatives $\delta'(\textbf{0})$ for expectation
values of some of the variables: ${\hat{g}_{0 \mu}}, 
{\hat{g}_{i j}}, {\hat{N}^{p}}$ and ${{\hat{\Gamma}}^{\alpha}_{~\mu \nu}}$. This is valid for all
physical states and is physically undesirable. 
Lastly, the above problems will arise if we use any set of metric compatible connections.

In the first approach to quantize the theory, it is unlikely that there
will exist a set of gauges that is time dependent, render the complete set of 
constraints second class and also remove the inconsistency mentioned above. It is
not possible to remove the inconsistency in the second approach to quantize
the theory. Also, it is expected that 
$[{\hat{g}_{\alpha \beta}}(x^{\mu}), {\hat{g}_{\alpha \beta}}(y^{\nu})]$ will depend on
$(x^{\mu}, y^{\nu})$ non-trivially with non-vanishing covariant derivatives [28]. 
We can consider semiclassical theories like
quantum fields in curved spaces to assume so. Thus, it is more appropriate to use
connections more general than metric compatible connections in quantum gravity.
Here, the constant time surfaces are imbedded in the four-manifold and 
the induced metric on the constant time surfaces coincide with the spatial metric $g_{i j}$. We can
use the metric compatible Levi-Civita connections associated with $g_{i j}$ on the constant time surfaces [26]
to demonstrate that this metric compatibility is not consistent with the quantization
condition given by Eqn.(19). The Levi-Civita connections of first kind associated with
$g_{i j}$ coincide with corresponding spatial parts of the Levi-Civita connections of first kind
of the four-manifold and it is more appropriate to introduce non-metricity in the
four-manifold. This is consistent with the discussions above. 
The above discussions are valid irrespective of the presence and nature of sources.
We can analyze this issue further in the following way. 
The Levi-Civita connections and metric compatibility conditions are taken as basic assumptions
to calculate the scalar curvature when we use the Einstein-Hilbert action to describe 
classical and quantum gravity. It is better to discuss quantization and non-metricity using the 
metric-affine action or Palatini action
where $C^{\alpha}_{~\mu \nu}$ in Eq.(9) is an independent field. 
In this article, non-metricity means ${{\nabla}_{\mu}}{g_{\alpha \beta}} \neq 0$.
A quantitative definition is given by Eq.(54).
We will discuss the corresponding variational problems and non-metricity in the next section.
We will address the issue of $(3 + 1)$ -decomposition of spacetime into space and time
in presence of non-metricity below Eq.(55) in section:VI.

\section*{IV. Affine Connections and the Lagrangian Formalisms}

In this section, we will discuss the kinematics of the metric-affine and Palatini formalisms.
This will help us to identify the degrees of freedom when
we use these formalisms that include affine connections as the independent
variables. This is important in the quantum theory
to impose quantization conditions which require non-metricity. 
In the Palatini formalism we only consider symmetric $C^{\alpha}_{~\mu \nu}$
in Eqs.(9,10). We also consider sources that do not couple with $C^{\alpha}_{~\mu \nu}$.
Important examples are source free theory and minimally coupled scalar and electromagnetic
field in the semiclassical limit of quantum gravity which presently is
a locally Lorentz invariant theory of quantum fields in curved spaces. 
Another important class of examples are many astrophysical systems like the Solar system.
In the metric-affine gravity we consider general $C^{\alpha}_{~\mu \nu}$
and also sources that couple with $C^{\alpha}_{~\mu \nu}$. An example is
fermions which couple with torsion through local Lorentz invariance in the semiclassical limit of
quantum gravity [41]. However, as
we have found in the previous section, we require non-metricity even in 
source-free quantum gravity. Thus, in the following we first consider the
solutions of the variational problem in the metric-affine gravity
with sources that do not couple with $C^{\alpha}_{~\mu \nu}$.
This can include the source-free theory and the Palatini formalism as special cases. We will 
consider coupling of matter field with $C^{\alpha}_{~\mu \nu}$ at the 
later part of the present section.

In the metric-affine theory, 
the Ricci tensor is given by: $R_{\mu \nu} = R_{\mu \kappa \nu}^{~~~~\kappa}$, where
$R_{\mu \kappa \nu}^{~~~~\kappa}$ can be obtained from Eq.(13). 
The Ricci tensor, scalar curvature and metric-affine action are given by the following 
expressions when we use affine connections given by Eqs.(9,10):

\ba
R_{\mu \alpha} & = & {{R'}_{\mu \alpha}} + 
2{{\nabla'}_{[\kappa}}{C^{\kappa}_{~\mu] \alpha}}
+ 2 [{C^{\lambda}_{~[\mu |\alpha|}} {C^{\kappa}_{~\kappa |\lambda|]}}] \\ \nonumber
R & = & {R'} + 2{g^{\mu \alpha}} \left\{{{\nabla'}_{[\kappa}}{C^{\kappa}_{~\mu] \alpha}}
+ [{C^{\lambda}_{~[\mu |\alpha|}} {C^{\kappa}_{~\kappa |\lambda|]}}]\right\} \\ \nonumber
S & = & {\int}{\sqrt{-g}}{R}{\bf e} + {\kappa_{M}}{S_M}(\psi, g_{\mu \nu})
\ea

\noindent{Where, ${{R'}_{\mu \alpha}}$ is the Ricci tensor evaluated using the 
Levi-Civita connections and $R'$ is the corresponding scalar curvature.
$\psi$ is matter field and the matter field action ${S_M}(\psi, g_{\mu \nu})$ does not
contain $C^{\alpha}_{~\mu \nu}$. $\kappa_{M}$ is a constant depending
on the nature of source [28]. In this article, by matter fields we will mean
both matter and gauge fields unless otherwise stated.  
We now extremize the action given by the last 
equation \textit{w.r.t} $C^{\alpha}_{~\mu \nu}$.
Covariant derivatives of the scalar density ${\sqrt{-g}}$ are
zero when we use the Levi-Civita connections.
The second term of the scalar curvature gives a boundary term by 
the Levi-Civita connections of ${{\nabla'}_{\mu}}$ and
Gauss's law form of Stoke's theorem [26,28]. This boundary term vanishes when 
$C^{\alpha}_{~\mu \nu}$ is held fixed at the boundary.
We then have the following equation as 
the solution of variational problem when $C^{\alpha}_{~\mu \nu}$ is held fixed
at the boundary:}

\be
{C^{\kappa}_{~\kappa \lambda}}{g^{\mu \alpha}} +
{C^{\alpha \kappa}_{~~~\kappa}}{{\delta^{\mu}_{\lambda}}} 
- {C^{\mu \alpha}_{~~~\lambda}} - {C^{\alpha ~ \mu}_{~\lambda}} = 0
\ee

\noindent{There is no contribution from the source fields in the present case. 
This is an algebraic equation giving constraints on $C^{\alpha}_{~\mu \nu}$.
We obtain the following equation when we extremize the action
\textit{w.r.t} $g_{\mu \nu}$:}

\ba
{\mathcal{G}}_{(\mu \alpha)} & = & 
{\mathcal{R}}_{(\mu \alpha)} - {1 \over 2}{\mathcal R}{g_{\mu \alpha}} = {8 \pi}{P_{\mu \alpha}} \\ \nonumber
{\mathcal{R}}_{\mu \alpha} & = & {{R'}_{\mu \alpha}} + 2 
[{C^{\lambda}_{~[\mu |\alpha|}} {C^{\kappa}_{~\kappa |\lambda|]}}] \\ \nonumber
{\mathcal R} & = & {g^{\mu \alpha}}{{\mathcal{R}}_{\mu \alpha}} 
\ea

\noindent{Where, ${\mathcal{R}}_{\mu \nu}$ and ${\mathcal{G}}_{\mu \nu}$ are the 
modified Ricci tensor and modified Einstein tensor respectively. They coincide with the
Ricci tensor and Einstein tensor when ${C^{\alpha}_{~\mu \nu}}$ is absent.
${P_{\mu \alpha}}$ is matter field stress tensor which is related to the variational 
derivative of matter field action \textit{w.r.t} $g^{\mu \alpha}$ by:
${8 \pi}{P_{\mu \alpha}} = - {\kappa_{M} \over {\sqrt{-g}}}{{\delta{S_M}} \over {\delta{g^{\mu \alpha}}}}$.
Here $\delta$ denotes functional derivatives and ${P_{\mu \alpha}}$ is symmetric.
Again, the second term of the curvature scalar does not contribute to Einstein's equation by 
the Levi-Civita connections of ${{\nabla'}_{\mu}}$ and the
Gauss's law form of Stoke's theorem when metric is held fixed at the boundary. 
The first order change in ${R'_{\mu \alpha}}$ due to change in metric
is given by: ${{\nabla'}_{[\kappa}}{{\delta \Gamma}^{\kappa}_{~\mu] \alpha}}$,
where ${{\delta \Gamma}^{\kappa}_{~\mu \alpha}}$ is the change in Levi-Civita connection
due to change in metric. This term does not contribute to the
equation of motion by the Stoke's theorem. Here, we have assumed that
metric and its first order derivatives are held fixed at the boundary [28].

We now construct the solutions of Eq.(23). 
A contraction over $(\lambda,\mu)$ leads to the following equation:}

\be
{H^{\kappa ~ \alpha}_{~\kappa}} +
{3 \over 2}{{\tilde{C}}^{\alpha \kappa}_{~~~\kappa}} = 0
\ee

\noindent{An alternate contraction over $(\alpha,\mu)$ leads to the following equation:}

\be
4{H^{\kappa}_{~\kappa \lambda}} +
2{{\tilde{C}}^{\kappa}_{~\kappa \lambda}} +
{{\tilde{C}}^{~ \kappa}_{\lambda ~\kappa}} = 0 
\ee

\noindent{Here, ${{\tilde{C}}^{\alpha}_{~\mu \nu}}$ is the symmetric part of
${{C}^{\alpha}_{~\mu \nu}}$ and ${H}^{\alpha}_{~\mu \nu} = 
{1 \over 2}({C^{\alpha}_{~\mu \nu}} - {C^{\alpha}_{~\nu \mu}}) = {1 \over 2}{{T}^{\alpha}_{~\mu \nu}}$, is 
half of torsion tensor. These two equations give the following equations from Eq.(23):}

\be
{H^{\kappa}_{~\kappa \lambda}}{g_{\mu \alpha}} +
{H^{\kappa}_{~\kappa(\alpha}}{g_{\mu)\lambda}} +
3{\tilde{C}_{(\mu \alpha)\lambda}} = 0
\ee

\noindent{and}

\be
H^{\kappa}_{{~\kappa}[\alpha}{g_{\mu]\lambda}} +
3{H}_{[\mu \alpha]\lambda} = 0
\ee

\noindent{Eqs.(27,28) give a set of homogeneous equations for ${C^{\alpha}_{~\mu \nu}}$
that can not determine ${C^{\alpha}_{~\mu \nu}}$ uniquely. We have four-fold ambiguities
associated with the projective transformation given by:
${C^{\alpha}_{~\mu \nu}} \rightarrow {C^{\alpha}_{~\mu \nu}} + 
{\delta^{\alpha}_{~\nu}}{\xi_{\mu}}$, where ${\xi_{\mu}}$ is a regular covariant vector field [38]. 
This is a symmetry of the gravitational part of metric-affine action and also of Eq.(23).
Projective transformation can introduce a limited version of non-metricity.
This no longer remains valid for the Palatini formalism where we only consider
symmetric ${C^{\alpha}_{~\mu \nu}}$. Eqs.(27,28) give us the Levi-Civita connections when we  
set ${H^{\kappa}_{~\kappa \mu}} = 0$ which fixes ${\xi_{\mu}} = 0$.
Projective invariance need not to be pertinent when we add sources in the metric-affine theory.
Without any dynamic equation for itself, ${\xi_{\mu}}$
can not produce non-trivial vacuum effects. Thus we will no longer consider projective
transformation in the following. We now consider the case when ${C^{\alpha}_{~\mu \nu}}$
is purely antisymmetric in the lower indices. In this case, we have the following
equations:}

\be
{H}_{[\mu \alpha] \lambda} = 0 ~;
\ee

\noindent{This gives vanishing solutions. With purely symmetric connections in the Palatini formalism, 
Eq.(27) also gives vanishing solutions.}

The above formalism will give non-vanishing solutions for $C^{\alpha}_{~\mu \nu}$
in presence of sources that couple with $C^{\alpha}_{~\mu \nu}$.
This is the case in the metric-affine gravity with fermions if
we consider locally Lorentz invariant theory
of quantum fields in curved spaces. Eq.(23) is replaced by the following
expression:

\ba
{C^{\kappa}_{~\kappa \lambda}}{g^{\mu \alpha}} +
{C^{\alpha \kappa}_{~~~\kappa}}{{\delta^{\mu}_{\lambda}}} 
- {C^{\mu \alpha}_{~~~\lambda}} - {C^{\alpha ~ \mu}_{~\lambda}} 
& = & {\Delta_{\lambda}^{~ \mu \alpha}} \\ \nonumber
{\Delta_{\lambda}^{~ \mu \alpha}} & = & 
- {\kappa_{M}}{{{\delta{S_M}}(\psi, g_{\mu \nu}, {C^{\lambda}_{~\mu \nu}})}
\over{\delta{C^{\lambda}_{~\mu \nu}}}} 
\ea

\noindent{Where, ${\Delta_{\lambda}^{~ \mu \alpha}}$ is known as the 
hypermomentum [32,38].
However, matter field Lagrangians are usually polynomials 
in first order covariant derivatives. In addition, gauge fields do not 
couple with ${C^{\alpha}_{~\mu \nu}}$ to express the corresponding field-strength tensors
as gauge-covariant curls [41]. We will illustrate this briefly in section:VI.
Thus, the above equation remains an algebraic constraint. Hence, it is appropriate to
extend the metric-affine and Palatini formalisms in quantum gravity to have nontrivial dynamics
for ${C^{\alpha}_{~\mu \nu}}$. In the next section, we will discuss a few theories as possible candidates
for a theory of gravity with dynamic ${C^{\lambda}_{~\mu \nu}}$.}

In the metric-affine theory with sources, we have the scope 
to have finite ${{\tilde{C}}^{\alpha}_{~\mu \nu}}$
when torsion is finite unless we have a special source such that: ${H^{\kappa}_{~\kappa \lambda}} = 0$. 
The antisymmetric part of the Ricci tensor is given 
by the following expression:

\be
R_{[\mu \alpha]} = {{\nabla'}_{\kappa}}{H^{\kappa}_{~\mu \alpha}} -
{{\nabla'}_{[\mu}}{C^{\kappa}_{~|\kappa| \alpha]}} +
{H^{\lambda}_{~\mu \alpha}}{C^{\kappa}_{~\kappa \lambda}} +
\left({\tilde{C}^{\lambda}_{~\kappa \alpha}}{H^{\kappa}_{~\lambda \mu}} -
{\tilde{C}^{\lambda}_{~\kappa \mu}}{H^{\kappa}_{~\lambda \alpha}}\right) \\ \nonumber
\ee

\noindent{Where, ${{\nabla'}_{\kappa}}$ is evaluated using the Levi-Civita
part of complete connections. Note that, $R'_{\mu \alpha}$ is symmetric because:
${{\Gamma}^{\kappa}_{~\kappa \alpha}} = {{\partial_{\alpha}}(ln{\sqrt|g|})}$,
where $|g|$ is the absolute value of the metric determinant.
Thus, in the metric-affine formalism, the Ricci tensor is not symmetric in general in
presence of sources. A purely symmetric Ricci tensor will impose $6$ additional 
constraints on the sources. The same is valid for the modified Ricci tensor. In
this case, the derivative terms will be absent.
This is also important to construct a semiclassical theory of fermions 
in a curved spacetime that can be derived from a variational principle.}

The following discussions in this paragraph are valid when $(3 + 1)$ -splitting
is possible. This is a non-trivial issue when we use general affine connection
and will be discussed below Eq.(55). 
In the metric-affine formalism discussed above, we can quantize the theory 
by considering $({g_{\mu \nu}}, {C^{\alpha}_{~\mu \nu}})$ as the complete set
of variables. Conjugate momenta of ${C^{\alpha}_{~\mu \nu}}$ 
vanish if we discard the four-divergences in the Lagrangian density. 
Affine connections do not spoil the diffeomorphism 
invariance of the theory and the constraints will include three sets: 
(i) four secondary constraints
coming from the non-dynamical nature of the lapse and shift functions and four
gauge fixing conditions; 
(ii) $64$ primary constrains: ${\pi_{\alpha}^{~\mu \nu}} = 0$;
(iii) $64$ secondary constraints given by Eq.(23) which are equivalent
to: ${C^{\alpha}_{~\mu \nu}} = 0$.
The complete set of constraints are second class with
suitable gauge fixing conditions and we can
try to impose quantization conditions similar to Eq.(19).
We will again have inconsistencies similar to that discussed in the
previous section if we use the Levi-Civita connections. 
With ${C^{\alpha}_{~\mu \nu}} = 0$ being the unique
solutions of Eq.(23), we will have to extend the standard Palatini formalism
to quantize generalized free gravity. This will also lead to general
affine connections. These aspects are also consistent with the discussions given below Eq.(29).

We conclude this section with a few comments on commutators.  
We consider the field-field commutators for independent variables at two general space time points:

\be
[{{\hat{\phi}}_{M}}(x^{\mu}),{{\hat{\phi}}_{M}}(y^{\nu})] = 
i{D_{M,M}}(x^{\mu},y^{\nu},{{\hat{\phi}}_{P}})
\ee

\noindent{Where, ${\hat{\phi}}_{M}$ is an independent field operator from the 
complete set of fields that include metric and ${C^{\alpha}_{{~\mu \nu}}}$.
We will use the mixed tensor ${C^{\alpha}_{{~\mu \nu}}}$
to find the commutators since it can give covariant derivatives of mixed tensors.  
${D_{M,M}}$ will depend on the type of field and should be consistent with
connection coefficients of the theory. In the case of gravity, general commutators: 
$[{\hat{g}_{\alpha \beta}}(x^{\mu}), {\hat{g}_{\alpha \beta}}(y^{\nu})]$ will depend on
$(x^{\mu}, y^{\nu})$ nontrivially, and it is appropriate to use
connections more general than metric compatible connections in the quantum theory.  
These metric-metric commutators will satisfy equations of
the form:}

\be
{{\hat{\nabla}}_{x \tau}}{D_{\alpha \beta, \alpha \beta}} = 
2 i [{\hat{C}_{(\alpha |\tau| \beta)}}(x^\mu), {\hat{g}_{\alpha \beta}}(y^{\nu})]
\ee

\noindent{The equal time commutators are supported at $\vec{x} = \vec{y}$ in locally or globally
Lorentz invariant quantum field theories. This may not remain strictly valid in
quantum gravity, in particular when (3 + 1) -decomposition is not possible.}

\section*{V. Affine Connections and the Extended Lagrangian Formalisms}

The variational problems discussed in the previous section give two sectors of
equations. The first sector is obtained by varying the action \textit{w.r.t}
${{C}^{\alpha}_{{~\mu \nu}}}$ and is given by Eqs.(23,30). 
Eq.(23) is homogeneous and contains no time derivatives. 
It acts like a constraint. We obtain Eq.(30) when we include suitable sources.
However, Eq.(30) is also an algebraic equation for the known sources which
are presently fermions coupled with torsion and we will have to extend these
formalisms to introduce dynamic ${{C}^{\alpha}_{{~\mu \nu}}}$. 
We will also have to extend the metric-affine formalism developed
for quantum field theory in curved spaces where metricity is a constraint to have locally
Minkowskian structure of the background spacetime [38,41].
We first try to extend the formalism to construct 
a quantum theory with finite and dynamic ${{C}^{\alpha}_{{~\mu \nu}}}$.
Here, we extend the formalisms in two alternate ways.   
The equations obtained are differential equations with time derivatives.

We can construct a theory by using the potential formalism.
By potential formalism we mean a formalism where ${C^{\alpha}_{~\mu \nu}}$
is derived from a tensor of lower rank. This is case with the Levi-Civita
connections that are derived from metric.
We will then have a set of differential equations 
including time derivatives in place of Eqs.(23,30) 
and we can introduce non-metricity even in the source-free theory by using nontrivial
solutions of the homogeneous differential equations. 
We can use Eqs.(11,12) to define symmetric connections, where 
${a_{\mu \nu}}, {b_{\mu \nu}}$ are symmetric. Non-metricity
is ensured by: $a_{\mu \nu} \neq k{g_{\mu \nu}}$, where $k$ is a 
constant which can be zero. This is a mild constraint, the otherwise
of which gives the Levi-civita connections. The Ricci tensor is symmetric in this case. 
This is obvious from ${{\nabla'}_{\mu}} \equiv {{\partial}_{\mu}}$ and Eq.(31).
We have:} 

\be
{{C}^{\kappa}_{~\kappa \alpha}} = {{S}^{\kappa}_{~\kappa \alpha}} = {{\partial_{\alpha}}(ln{\sqrt|b|})};
{~} {{H}^{\kappa}_{~\alpha \beta}} = 0
\ee

\noindent{Where $b$ is the determinant of $b_{\mu \nu}$ given in Eqs.(11,12).
We have: ${{\partial_{[\mu}}{\partial_{\alpha]}}(ln{\sqrt|b|})} = 0$ and 
$R_{[\mu \nu]}$ vanishes by Eq.(31). Thus, it is appropriate to use $a_{\mu \nu}$
to describe the connections when the Ricci tensor is symmetric. 
Note that ${\sqrt|b|}$ is a scalar density. In the next section, we will consider two simple cases that
give non-metricity and that can be useful to explain cosmological accelerations.}

We now discuss another model with zero torsion. Thus, we are
considering the Palatini formalism.
We asuume that ${{\tilde{C}}^{\alpha}_{~\mu \nu}}$ is given by:

\be
{\tilde{C}}^{\alpha}_{~\mu \nu} = {{\nabla'}^{\alpha}}{q_{\mu \nu}}
\ee

\noindent{Here ${q_{\mu \nu}}$ is a symmetric tensor and ${{\nabla'}_{\alpha}}$
uses the Levi-Civita connections given by Eqs.(9,10). This can be useful when
the Ricci tensor is not symmetric. We put this expression
in the Palatini action given by Eqs.(8,22). We then take variational derivatives \textit{w.r.t} 
${q_{\mu \nu}}$. In this case, the action contains a second order derivative which does not
contribute to the variational derivative by the Levi-Civita connections of
${{\nabla'}_{\mu}}$ and the Gauss's law form of Stoke's theorem [26]. 
We have the following equation:

\be
2{{\nabla'}^{\kappa}}{{\nabla'}^{(\mu}}{q^{\alpha )}_{~\kappa}}
- {g^{\mu \alpha}}{{\nabla'}_{\lambda}}{{\nabla'}^{\kappa}}{q^{\lambda}_{~\kappa}}
- {{\nabla'}^{\mu}}{{\nabla'}^{\alpha}}{q} = 0
\ee

\noindent{Where, $g_{\mu \nu}$ is a solution of the
metric sector Einstein's equation,
${{\nabla'}_{\lambda}}$ is the corresponding metric compatible 
covariant derivative that uses the Levi-Civita connections 
and $q$ is the trace of ${q_{\mu \nu}}$.
All the above equations are dynamical and all conjugate momenta
are finite. Moreover, time derivative of all
components of ${q^{\lambda}_{~\kappa}}$ are present in the complete
set of equations. In this case, the modified Ricci tensor is symmetric and 
the Ricci tensor can be made symmetric by
imposing the constraint: ${{\nabla'}_{[\mu}}{\tilde{C}}^{\kappa}_{~|\kappa| \alpha]} = 0$.
This leads us to introduce a scalar field:

\be
{{\nabla'}^{\kappa}}{q_{\kappa \alpha}} = - {{\nabla'}_{\alpha}}{\zeta}
\ee

\noindent{Eq.(36) gives us the following relation between the trace of $q_{\mu \nu}$ and $\zeta$:}

\be
{{\nabla'}_{\mu}}{{\nabla'}^{\mu}}(q) = 2 {{\nabla'}_{\mu}}{{\nabla'}^{\mu}}(\zeta)
\ee

\noindent{There will be a non-holonomic constraint to ensure nonmetricity: 
${{\nabla'}^{(\mu}}{q^{\alpha )}_{~\kappa}} \neq 0$. The trace of 
${{\nabla'}^{(\mu}}{q^{\alpha )}_{~\kappa}}$ is $- {{\nabla'}_{\kappa}}{\zeta}$,
when the corresponding Ricci tensor is symmetric.}

We can also construct a special model with:
${H}^{\alpha}_{~\mu \nu} = {{\nabla'}^{\alpha}}{f_{\mu \nu}}$, 
where ${f_{\mu \nu}}$ is an antisymmetric tensor, and include both $q_{\mu \nu}$ and $f_{\mu \nu}$.
A detail potential formalism of torsion is discussed in [42].
Here we have the following set of equations for $q_{\mu \nu}$ and $f_{\mu \nu}$:}

\ba
2{{\nabla'}^{\kappa}}{{\nabla'}^{(\mu}}{q^{\alpha )}_{~\kappa}}
- {g^{\mu \alpha}}{{\nabla'}_{\lambda}}{{\nabla'}^{\kappa}}{q^{\lambda}_{~\kappa}}
- {{\nabla'}^{(\mu}}{{\nabla'}^{\alpha)}}{q} + 
{g^{\mu \alpha}}{{\nabla'}^{\lambda}}{{\nabla'}^{\kappa}}{f_{\lambda \kappa}} 
& = & 0 \\ \nonumber
{{\nabla'}^{\lambda}}\left\{{{\nabla'}^{[\mu}}{{f^{\alpha]}_{~ \lambda}}}\right\}
- {{\nabla'}^{[\mu}}{{\nabla'}^{\alpha]}}{q} & = & 0
\ea

\noindent{This is a set of coupled differential equations which are
dynamical in nature. The last term in both the equations
give coupling between $q_{\mu \nu}$ and $f_{\mu \nu}$ and vanishes when torsion
is vanishing. We again find that in the metric-affine theory we have finite
$q_{\mu \nu}$ when $f_{\mu \nu}$ is finite. 
Non-metricity need not to preserve local Minkowskian structure
of spacetime [38], and the particle interpretations of these fields 
are not obvious. These fields, although massless, can be important in
inflation driven by scalar as well as higher spin fields 
[43,44,45,46]. We will later discuss the dynamics of this model.
If required, we can also use other potentials for torsion [42].}

Alternatively, we can use different actions to
construct the dynamics of ${{C}^{\alpha}_{{~\mu \nu}}}$ field itself.
We can include higher order curvature scalars like ${R^{\mu \nu}}{R_{\mu \nu}}$ 
in the Einstein-Hilbert Lagrangian with the Riemann curvature tensor 
evaluated using finite ${C^{\alpha}_{~\mu \nu}}$ as given by Eq.(13). 
This is interesting if we consider the renormalizability of quantum gravity.
We can also use extended theories of gravity like 
the Palatini $f(R)$ gravity or metric-affine $f(R)$ gravity.
One can look at [6,8] for reviews on this topic. 
When we use the curvature scalar given by Eq.(22) in a metric-affine
$f(R)$ gravity, we can no longer neglect the total divergence middle term:
$2{g^{\mu \alpha}}{{\nabla'}_{[\kappa}}{C^{\kappa}_{~\mu] \alpha}}$. It will
couple with other terms of $R_{\mu \nu}$ and we will have differential equations 
for ${C^{\kappa}_{~\mu \alpha}}$  in general.
This can give dynamic on shell ${C^{\kappa}_{~\mu \alpha}}$. We will have to ensure non-metricity. 
We will again have finite ${{\tilde{C}}^{\alpha}_{{~\mu \nu}}}$ in general when
${{H}^{\alpha}_{{~\mu \nu}}}$ is finite. 
We can also use the potential formalism in such theories.
A relevant formulation is the dynamical theory of metric 
compatible torsion (contorsion) discussed by a few 
authors with different actions [47,48]. These papers introduce metric compatible
torsion in an effective action of quantum gravity obtained from the
Einstein-Hilbert action and discuss propagators for the corresponding particles. 
Corresponding potential theory of torsion is ghost free in this formalism [47].
For our purpose, it is appropriate to extend the Einstein-Hilbert-Palatini
action at the classical level. We can also use theories like string theory.}

\section*{VI. Applications in Cosmology}

In this section, we will consider applications of the potential
formalism of the Palatini theory discussed previously to
introduce finite on shell non-metricity. 
Corresponding ${{C}^{\alpha}_{{~\mu \nu}}}$ are symmetric in the lower indices
and the sources do not couple with these fields. We have mentioned some examples
in sections:II,V. We also consider symmetric
Ricci tensors only. This is the most relevant case when we consider semiclassical and
classical limits of quantum gravity. This example will introduce two massless 
scalar fields. One of them will contribute a negative stress tensor to Einstein's equation
in the semiclassical or classical theories.
We will use the geometrized units in the following where, $G = c = 1$.

In section:II, we have mentioned that we
can introduce non-metricity by defining the affine connections to be compatible
with a symmetric covariant field: ${b_{\mu \nu}} = {g_{\mu \nu}} + {a_{\mu \nu}};
{~} {a_{\mu \nu}} \neq k{g_{\mu \nu}}$, where $k$ is a constant including zero.  
In the last section we have found that such a set of affine connections 
give a symmetric Ricci tensor. We now break ${a_{\mu \nu}}$ into a trace and a
traceless part:

\be
{a_{\mu \nu}(x)} = {\Phi}(x){g_{\mu \nu}} + {{\bar{a}}_{\mu \nu}(x)}; 
{~~} {{\Phi}(x)} = {a(x) \over 4} 
\ee

\noindent{Where, $\Phi$ is a scalar field, $a(x)$ is the trace 
of ${a_{\mu \nu}}$ and ${{\bar{a}}_{\mu \nu}}$ is trace-free.
We can express corresponding ${{\tilde{C}}^{\alpha}_{~\mu \nu}}$ in the following
way:}

\be
{{\tilde{C}}^{\alpha}_{~\mu \nu}} = {{\delta}^{\alpha}_{~ (\mu}}{{\nabla'}_{\nu)}}{[ln(1 + \Phi)]}
- {1 \over 2}{g_{\mu \nu}}{{{\nabla'}^{\alpha}}{[ln(1 + \Phi)]}} + {D^{\alpha}_{~~\mu \nu}}
\ee

\noindent{The first two terms in the \textit{r.h.s} gives the contribution of the trace
part of $b_{\mu \nu}$ given by: $({g_{\mu \nu}} + {\Phi(x)}{g_{\mu \nu}})$, [28]. 
${D^{\alpha}_{~~\mu \nu}}$ can be expressed
in terms of ${g_{\mu \nu}}, {a_{\mu \nu}}$
and their derivatives. A further
contraction of the third term in the covaraiant indices leads to the following expression
for a general $a_{\mu \nu}$:

\ba
{{\tilde{C}}^{\alpha}_{~\mu \nu}} & = & {{\delta}^{\alpha}_{~ (\mu}}{{\nabla'}_{\nu)}}{[ln(1 + \Phi)]}
- {1 \over 2}{g_{\mu \nu}}{{{\nabla'}^{\alpha}}{[ln(1 + \Phi)]}}  
+ {g_{\mu \nu}}{H^{\alpha}} + {{E}^{\alpha}_{~~\mu \nu}} \\ \nonumber
{~} & = & {{\delta}^{\alpha}_{~ (\mu}}{{\nabla'}_{\nu)}}{[ln(1 + \Phi)]}
- {1 \over 2}{g_{\mu \nu}}{{{\nabla'}^{\alpha}}{[ln(1 + \Phi)]}}  
+ {g_{\mu \nu}}{{\nabla'}^{\alpha} {\Psi}} + {g_{\mu \nu}}{B^{\alpha}} + {E^{\alpha}_{~~\mu \nu}}
\ea

\noindent{where $\Phi$ gives the trace-scalar of $a_{\mu \nu}$ and ${{E}^{\alpha}_{~~\mu \nu}}$ is traceless in the lower indices.
$\Psi$ is a scalar field with ${{\nabla'}^2}{\Psi} = {{\nabla'}_{\mu} {H^{\mu}}}$.
${B^{\alpha}}$ is a vector field with: ${{\nabla'}_{\alpha} {B^{\alpha}}} = 0$, and is
suitable to describe a spin one boson. However, we also have: ${{\nabla'}_{[ \mu} {B_{\nu ]}}} = 0$, 
which follows from the symmetry of the Ricci tensor that remains valid with a symmetric
$a_{\mu \nu}$. In the following we will take $B^{\mu} = 0$. $\Psi$ is expected. 
In the Minkowski space, we have a spin zero boson associated
with ${{\bar{a}}_{\mu \nu}}$ in addition to a spin one boson and a spin two boson.
We will later briefly discuss ${E^{\alpha}_{~~\mu \nu}}$ which is expected to describe 
a spin two boson.}

We now consider the simplest cases of Eq.(42). We first consider
the case where the traceless part of $a_{\mu \nu}$ vanish
and which introduces a scalar field only:

\be
{b_{\mu \nu}} = {g_{\mu \nu}} + {\Phi(x)}{g_{\mu \nu}}
= {\chi(x)}{g_{\mu \nu}}
\ee

\noindent{We use affine connections that are compatible with ${b_{\mu \nu}}$
which is conformal to metric. In this case, ${\tilde{C}}^{\alpha}_{~\mu \nu}$ 
and the modified curvature scalar are given by the following expressions [28]:}

\ba
{{\tilde{C}}^{\alpha}_{~\mu \nu}} & = & {{\delta}^{\alpha}_{~ (\mu}}{{\nabla'}_{\nu)}}{[ln(1 + \Phi)]}
- {1 \over 2}{g_{\mu \nu}}{{{\nabla'}^{\alpha}}{[ln(1 + \Phi)]}} \\ \nonumber
{\mathcal{R}} & = & {R'}  - {3 \over 2} {1 \over (1 + \Phi)^{2}}{[{\nabla'}(1 + {\Phi})]^{2}} 
= {R'}  - {3 \over 2}{[{\nabla'}{ln(\chi)}]^{2}}
\ea

\noindent{Here ${({\nabla'}{\Phi})^{2}} = ({{\nabla'}_{\mu}}{\Phi})({{\nabla'}^{\mu}}{\Phi})$, 
is the norm of the gradient of $\Phi$ and the primed quantities are evaluated using the Levi-Civita
connections. For small $\Phi ~(<< 1)$, we have the following modification
of the \textit{r.h.s} of Einstein's equation where covariant derivatives
are evaluated using the Levi-Civita connections [28]:}

\ba
{1 \over {\chi}}{{{\nabla'}_{\kappa}}{{\nabla'}^{\kappa}}{ln(\chi)}} & = &
{1 \over {\chi^2}}[{{{\nabla'}_{\kappa}}{{\nabla'}^{\kappa}}{\chi}} -
{1\over {\chi}}{({\nabla'}{\chi})^{2}}] = 0 {~~} {\approx} {~~}
{{{\nabla'}_{\kappa}}{{\nabla'}^{\kappa}}{\Phi}} -
{({\nabla'}{\Phi})^{2}} \\ \nonumber
{{G'}_{\mu \alpha}} & = & {8 \pi}{{P'}_{\mu \alpha}}  
+ {3 \over {2 {\chi}^2}}[({{\nabla'}_{\mu}}{\chi})({{\nabla'}_{\alpha}}{\chi})
- {1 \over 2}{g_{\mu \alpha}}{({\nabla'}{\chi})^{2}}] \\ \nonumber
{~} & \approx & {8 \pi}{{P'}_{\mu \alpha}}   
+ {3 \over 2}[({{\nabla'}_{\mu}}{\Phi})({{\nabla'}_{\alpha}}{\Phi})
- {1 \over 2}{g_{\mu \alpha}}{({\nabla'}{\Phi})^{2}}] \\ \nonumber
{~} & = & {8 \pi}[{{P'}_{\mu \alpha}} + {3 \over {16 \pi}}{{P'}_{\mu \alpha}(\Phi)}]
\ea

\noindent{Where, we have considered terms upto second order in $\Phi$. 
${{P'}_{\mu \alpha}}$ is the stress tensor of ordinary matter.
${{P'}_{\mu \alpha}}(\Phi)$ is the stress tensor of an ordinary masless
scalar field. Both $\chi$ and $\Phi$
are massless. The exact equations are consistent with: 
${{\nabla'}^{\alpha}}{{G'}_{\mu \alpha}} = 0$. 
We find that the scalar field $\Phi$ behaves like a 
massless scalar field with its stress tensor coming as a source term in
Einsrein's equation. This field can be useful to explain inflation. 
We can modify the metric-affine action by adding potential
term for $\chi$ or $\Phi$. For non-metricity, we have the following condition:}

\be
{{\nabla}_{\mu}{g_{\alpha \beta}}} = 
- {g_{\alpha \beta}}{{\nabla'}_{\mu}{[ln(1 + \Phi)]}} \neq 0 
\ee

\noindent{We will later discuss the significance. In the present case, matter fields do not
couple with $\Phi$. The only observable effect of ${\Phi}$ is
to produce a massless scalar field stress tensor in Einstein's equation.}

We now consider the other case where only $\Psi$ is finite:

\be
{\tilde{C}}^{\alpha}_{~\mu \nu} = {g_{\mu \nu}}{{\nabla'}^{\alpha} {\Psi}} 
\ee

\noindent{We have the following expressions for different quantities:}

\ba
{\mathcal{R}}_{\mu \alpha} & = & {{R'}_{\mu \alpha}} + 
{g_{\mu \alpha}}{({\nabla'}{\Psi})^{2}}
- ({{\nabla'}_{\mu}}{\Psi})({{\nabla'}_{\alpha}}{\Psi}) \\ \nonumber
{\mathcal{R}} & = & {R'} + 3 {({\nabla'}{\Psi})^{2}} 
\ea

\noindent{We now solve the corresponding extremization problem and
obtain the following set of equations:}

\ba
{{{\nabla'}_{\kappa}}{{\nabla'}^{\kappa}}{\Psi}} & = & 0 \\ \nonumber
{{G'}_{\mu \alpha}} & = & {8 \pi}{{P'}_{\mu \alpha}} - 3[({{\nabla'}_{\mu}}{\Psi})({{\nabla'}_{\alpha}}{\Psi})
- {1 \over 2}{g_{\mu \alpha}}{({\nabla'}{\Psi})^{2}}] \\ \nonumber
{~} & = & {8 \pi}[{{P'}_{\mu \alpha}} - {3 \over {8 \pi}}{{P'}_{\mu \alpha}(\Psi)}]
\ea

\noindent{Where, ${{P'}_{\mu \alpha}}$ is the stress-tensor of ordinary sources
and ${{P'}_{\mu \alpha}(\Psi)}$ is the stress-tensor of an ordinary massless scalar field. 
This stress tensor satisfy all the energy and pressure conditions that the stress-tensor of a massless
scalar field satisfies but it comes with an opposite sign in the \textit{r.h.s}
of Einstein's equation, when we state the equation
using the Levi-Civita connections. Thus, the effect of $\Psi$ introduced
to generalize the Levi-Civita connections, is to contribute a negative
massless scalar field stress tensor to the sources of the Einstein equation obtained from
the Einstein-Hilbert action formalism. This gives us an alternate way to explain effects that
dark energy is proposed for. Negative stress-tensor is also important in Hoyle-Narlikar theory of
gravity and also in wormhole and warp drive [49,50]. In this case, we can not
generate the cosmological constant with ordinary massless scalar field for which the energy density is positive definite.
We have the following expression for non-metricity:}

\be
{{\nabla}_{\mu}{g_{\alpha \beta}}} = 
- 2{g_{\mu (\alpha}}{{\nabla'}_{\beta )}{\Psi}}
\ee

\noindent{We now consider the case when both $(\Psi, \Phi)$ are present. ${{\tilde{C}}^{\alpha}_{~\mu \nu}}$
is given by the following expression:}

\be
{{\tilde{C}}^{\alpha}_{~\mu \nu}} = {{\delta}^{\alpha}_{~ (\mu}}{{\nabla'}_{\nu)}}{[ln(1 + \Phi)]}
- {1 \over 2}{g_{\mu \nu}}{{{\nabla'}^{\alpha}}{[ln(1 + \Phi)]}} + 
{g_{\mu \nu}}{{\nabla'}^{\alpha} {\Psi}} 
\ee

\noindent{We obtain the following expression for the modified curvature scalar:}

\be
{\mathcal{R}} = {R'}  - {3 \over 2} {1 \over (1 + \Phi)^{2}}{({\nabla'}{\Phi})^{2}} 
+ 3{({\nabla'}{\Psi})^{2}} + {3 \over (1 + \Phi)}[({{\nabla'}_{\kappa}{\Phi}})({{\nabla'}^{\kappa}{\Psi}})]
\ee

\noindent{We have the following generalization of Einstein's equation:}

\ba
{{{\nabla'}_{\kappa}}{{\nabla'}^{\kappa}}(1 + \Phi)} & - & {1\over {(1 + \Phi)}}{[{\nabla'}(1 + {\Phi})]^{2}} = 0 \\ \nonumber
{{{\nabla'}_{\kappa}}{{\nabla'}^{\kappa}}{\Psi}} & = & 0 \\ \nonumber
{{G'}_{\mu \alpha}} & = & {8 \pi}[{{P'}_{\mu \alpha}} + {3 \over {16 \pi}}{{P'}_{\mu \alpha}(\Phi)}
- {3 \over {8 \pi}}{{P'}_{\mu \alpha}(\Psi)} - 
{3 \over {8 \pi}}{{P'}_{\mu \alpha}(\Psi, \Phi)}] \\ \nonumber
{{P'}_{\mu \alpha}(\Psi, \Phi)} & = & {1\over {(1 + \Phi)}}
[({{\nabla'}_{( \mu}}{\Psi})({{\nabla'}_{\alpha )}}{\Phi})
- {1 \over 2}{g_{\mu \alpha}}({{\nabla'}_{\kappa}{\Psi})({{\nabla'}^{\kappa}}{\Phi}})] \\ \nonumber
& \approx & ({{\nabla'}_{( \mu}}{\Psi})({{\nabla'}_{\alpha )}}{\Phi})
- {1 \over 2}{g_{\mu \alpha}}({{\nabla'}_{\kappa}{\Psi})({{\nabla'}^{\kappa}}{\Phi}})
\ea

\noindent{The first two equations are the equations for $\Phi$ and $\Psi$
and they remain same as Eqn.(45) and Eqn.(49) respectively.
We find that coupling of $\Psi$ with $\Phi$ in ${\mathcal{R}}$ gives another
contribution to source stress tensor which can be positive or negative. 
This is again important for dark energy research.
It is interesting to note that we can always have a set of $a_{\mu \nu}$  
for which only $\Phi$ is present in spacetime. 
On the other hand, Eq.(47) corresponds to the choice:
${\Phi}, {{E}^{\alpha}_{~~\mu \nu}} = 0$, 
in Eq.(42). It is expected that we will have non-trivial solutions of the above constraints
and Eq.(47) that are in general finite when $\Psi$ is so. 
We will discuss this aspect after Eqn.(57).
Otherwise, we can always have Eq.(47) by using the 
symmetric potentials introduced in the previous section. We can choose: 
$q_{\mu \nu} = {\Psi}(x){g_{\mu \nu}}$, in Eq.(35). This will give a symmetric Ricci tensor with
$\zeta = - {\Psi}$, in Eq.(37).}

We now briefly discuss the geometrical significance of $(\Psi, \Phi)$ and corresponding non-metricities
given by Eqs.(46,50). We define the the non-metricity tensor by:

\be
{Q_{\mu \alpha \beta}} = - {{\nabla}_{\mu}}{g_{\alpha \beta}}
= {g_{\alpha \beta}}{{\nabla'}_{\mu}{[ln(1 + \Phi)]}}; {~~}
2{g_{\mu (\alpha}}{{\nabla'}_{\beta )}{\Psi}}
\ee

\noindent{Both $\Psi$ and $\Phi$ can be present in the manifold 
in a general theory. We can split ${Q_{\mu \alpha \beta}}$ into a trace $Q_{\mu}$ 
and traceless part $\bar{Q}$} in the last two indices [38]:

\be
{Q_{\mu \alpha \beta}} = {Q_{\mu}}{g_{\alpha \beta}} + {{\bar{Q}}_{\mu \alpha \beta}};
\ee

\noindent{Both trace and traceless parts of ${Q_{\mu \alpha \beta}}$ 
are finite for $\Psi$. The same is valid in general for ${B^{\alpha}}, {{E}^{\alpha}_{~~\mu \nu}}$. 
Corresponding connections do not preserve the light cone under parallel transport and we no longer have the 
local Minkowski structure of spacetime [32,38].
Thus, we can not have exact $(3 + 1)$ -splitting of the underlying manifold
into space and time. We find that departure from local Minkowski geometry can give new fields
like dark matter and dark energy not found ordinarily. 
${{\bar{Q}}_{\mu \alpha \beta}}$ vanishes for $\Phi$.
A few examples of non-metric affine connections with vanishing ${\bar{Q}}$, that are 
local gauge theories for the Weyl group (Poincare group plus dilatation), 
are discussed in the references given at the footnote $24$ of [38].
We can have finite or vanishing torsion in such theories.
These theories preserve the light cone under parallel transport
and are locally Minkowski due to the reparameterization invariance of the form of the geodesic 
equation: $t^{\mu}{{\nabla}_{\mu} t^{\nu}} = {f}t^{\nu}$, where $f$ is a scalar 
function on the curve [28]. Here, we can use 
a parameterization so that: $f = {1 \over 2}{t^{\mu}}{{\nabla'}_{\mu}{[ln(1 + \Phi)]}}$. 
However, this may cause problem to time-orientation when $\Phi$ is strong.}

The fields $\Phi, {\Psi}, q_{\mu \nu}$ are purely quantum gravitational in origin. 
They are kinematically required in quantum gravity. 
In this domain, ${Q_{\mu \alpha \beta}}$ can not vanish
due to the arguments given in Sect.III. 
$(\Phi, {\Psi})$ can be finite everywhere and hence, they are non-localized
similar to dark energy and inflation.
${\Psi}, q_{\mu \nu}$ break the local Minkowski structure of spacetime, 
and are not observed to be present in the matter-gravity coupling part of the semiclassical theory of quantum gravity
which presently is locally Lorentz invariant quantum field theory in curved spaces. 
It is unexpected that any large
coupling between ${\Phi}, {\Psi}, q_{\mu \nu}$ 
with ordinary matter like fermions, present
in the full quantum theory, will be lost in the semiclassical limit.
Thus, we assume that these fields can be present without 
corresponding sources from ordinary matter. More complete discussions can be found in
[51,52,53]. However, they can couple with dark matter. 
They contribute non-trivially to the source stress tensor of Einstein's equation
and are possible candidates for dark energy and inflation. We will illustrate this further in the following section.
Dark energy is non-localized, have negative pressure 
and is primarily observed by their gravitational effects. These make $\Psi$
a possible candidate for dark energy. $\Phi$ can be a possible candidate for inflation. 
The amount of dark energy is much higher than ordinary matter and
$\Psi$ need not to have ordinary matter with finite hypermomentum as its sources. 
A complete theory of quantum gravity will illuminate this issue further.
Regarding the scalar fields $({\Psi}, {\Phi})$, they are in line with other scalars introduced similarly, like the dilaton [54].
Observed local Lorentz invariance in many experiments indicate  
that $({\Psi}, {\Phi})$ are presently very small. This can explain the smallness
of parameters like the cosmological constant required to explain 
present cosmological acceleration in some models. With $({\Psi}, {\Phi})$
being small and non-localized, their effects are usually observed in large scale
phenomena and are not much significant in small scale astrophysics
like that of the Solar system. This is another characteristic aspect of dark energy.  
However, local large inhomogeneities and anisotropies in ${C^{\alpha}_{~~\mu \nu}}$ can
cause significant effects on geodesic motions in the corresponding regions.
Quantum fluctuations in $({\Psi}, {\Phi})$, including vaccuum fluctuations, can
also be useful to explain different cosmological eras.
When required, we can extend non-metricity and include torsion potentials to explain different cosmological observations. 
We can also modify the metric-affine action for this purpose. 
We will briefly discuss this issue at the end of this section.
A related problem is to find out possible interactions and particle interpretations 
of fields that correspond to different representations of the Lorentz group when the local
Minkowski structure of spacetime is broken strongly. This will happen
in the quantum gravitational domain. 
This will be important to understand the relations between the complete set of
fields including dark matter, dark energy, ordinary matter and gauge fields. We can continue to describe gauge
theories by potentials. This is consistent with the Palatini formalism, since the potentials are analogous
to connections in the geometric theory of gauge fields [55]. 
To illustrate, we can describe the electromagnetic field tensor 
by a set of potentials in a gauge invariant way, [38]:

\be
{F^{\alpha \beta}} = {g^{\alpha \mu}}{g^{\beta \nu}}{F_{\mu \nu}}\\ \nonumber
 = {g^{\alpha \mu}}{g^{\beta \nu}}
[{{\nabla_{\mu}}{A_{\nu}}} -  {{\nabla_{\nu}}{A_{\mu}}} 
+ {T^{\alpha}_{~\mu \nu}}{A_{\alpha}}]
= {g^{\alpha \mu}}{g^{\beta \nu}}[{{\partial}_{\mu}}A_{\nu} -  
{{\partial}_{\nu}}{A_{\mu}}]
\ee

\noindent{The above definition is valid in the metric-affine theory with finite torsion.
${T^{\alpha}_{~\mu \nu}}$ vanish in the Palatini formalism. We find that $A_{\mu}$ do not
couple with ${{C}^{\alpha}_{~\mu \nu}}$. This aspect remains valid
for the non-Abelian gauge theories where the field-strength tensors are given by
gauge-covariant curls of the corresponding potentials [56]. The discussions above Eq.(56) are new perspectives 
in quantum gravity deduced mathematically in this article and partly 
supported by cosmological observations.}

We can try to introduce coupling between $({\Psi}, {\Phi})$ and fermions 
in quantum gravity by removing the antisymmetry and metric compatibility conditions on contorsion 
provided the non-local character of $({\Psi}, {\Phi})$ is maintained. We will also have to explain
why such coupling is not significant in the semiclassical and classical theories.
Otherwise, we can have observable effects of non-metricity
in events like NS-NS, NS-WD and WD-WD mergers [57,58].

\section*{VII. Extension of Einstein's Theory of Gravity}

We can use classical theories and quantum field theory in curved spaces to find the effects  
of ${\Phi}$ and ${\Psi}$. This will be useful in cosmic epoch. We now compare the following action discussed
in the present article:

\ba
S & = & {\int}{\sqrt{-g}}{R}{\bf e} + {\sum}{\kappa_{M}}{S'_M}({\chi_a}, g_{\mu \nu})
= {\int}{\sqrt{-g}}{\mathcal{R}}{\bf e} + {\sum}{\kappa_{M}}{S'_M}({\chi_a}, g_{\mu \nu}) \\ \nonumber
{\mathcal{R}} & = & {g^{\mu \alpha}}{\mathcal{R}_{\mu \alpha}}
= {g^{\mu \alpha}} \left\lbrace  {{R'}_{\mu \alpha}} 
+ 2 [{C^{\lambda}_{~[\mu |\alpha|}} {C^{\kappa}_{~\kappa |\lambda|]}}] \right\rbrace  \\ \nonumber
& = & {R'} - {3 \over 2}{({\nabla'}{\Omega})^{2}} 
+ 3{({\nabla'}{\Psi})^{2}} + 3 [({{\nabla'}_{\kappa}{\Omega}})({{\nabla'}^{\kappa}{\Psi}})], ~\Omega = ln(1 + \Phi)
\ea

\noindent{where ${\bf e}$ is the coordinate volume element ${dx^0} \wedge {dx^1} \wedge {dx^2} \wedge {dx^3}$ [28], with the action of scalar-tensor theories [6]:}

\be
S = {\int {\bf e}}{\sqrt{-g}}{\big[}f(\phi, R') - \zeta(\phi){({\nabla'}{\phi})^{2}}{\big]}
+ {S'_{m}(\chi_a , {g_{\mu \nu}})} 
\ee

\noindent{where $\chi_a$ is the collection of ordinary matter and radiation. Eq.(54) indicates that: $\zeta(\omega) = {3 \over 2}, ~~ \zeta(\psi) = - 3$. $f(\phi, R) = R$ for both fields. Thus, the scalars mentioned at the beginning of the first section like quintessence and k-essence can be related to quantum gravity and can have geometrical origin. When required, we can modify the theory by adding suitable $\omega$ and $\psi$ dependent terms in Eq.(54). We get:}

\be
\mathcal{L} = {\sqrt{-g}} (\mathcal{R} - \mathcal{V})
\ee

\noindent{where $\mathcal{V}$ contains the added terms which should be consistent with: ${{\nabla'}^{\alpha}}{{G}_{\mu \alpha}} = 0,$ if we express the equations in the form similar to Eq.(51). In general, the above Lagrangian density will give derivatively coupled equations for $\omega$ and $\psi$. We obtain a decoupled theory by diagonalizing the symmetric bilinear form in ${{\nabla'}^{\mu} {\Omega}}$ and ${{\nabla'}^{\nu} {\Psi}}$ given by Eq.(54), [51]. This bilinear form is similar to: ${a^2} - 2ab - 2{b^2}$ and the corresponding symmetric matrix has eigenvalues: ${\lambda_{\pm}} = (-1 \pm \sqrt{13})/2$ . $\mathcal{R}$ is now given by the following expression:

		\ba
		{\mathcal{R}} & = & {R'}  - {3 \over 2}[{({\nabla'}{\Omega})^{2}} 
		- 2{({\nabla'}{\Psi})^{2}} - 2 ({{\nabla'}_{\kappa}{\Omega}})({{\nabla'}^{\kappa}{\Psi}})] \\ \nonumber
		& = & {R}  - {3 \over 2}{g^{\mu \nu}}[{\lambda_{+}}{({{\nabla'}_{\mu}{\Phi'}})({{\nabla'}_{\nu}{\Phi'}})}
		+ {\lambda_{-}}{({{\nabla'}_{\mu}{\Psi'}})({{\nabla'}_{\nu}{\Psi'}})}] 
		\ea

		\noindent{where $\Phi'$ and $\Psi'$ are related to $\Omega$ and $\Psi$ by the following expression:}

		\be
		{\begin{pmatrix}
				\Phi' \\
				\Psi' 
		\end{pmatrix}}  = 
		{S^T} {\begin{pmatrix}
				\Omega \\
				\Psi 
		\end{pmatrix}},  ~~~ S = {\begin{pmatrix}
				{c^{+}_{1}} & {c^{-}_{1}} \\
				{c^{+}_{2}} & {c^{-}_{2}} 
		\end{pmatrix}}.
		\ee

		\noindent{Here ${c^{\pm}_{i}}$ are the components of two orthonormal eigenvectors. $S$ is given by the following expression:}

		\be
		S = 
		{\begin{pmatrix}
				{\sqrt{4 \over {26 - 6\sqrt{13}}}} &  {\sqrt{4 \over {26 + 6\sqrt{13}}}} \\
				{{3 - \sqrt{13} \over 2}}\sqrt{4 \over {26 - 6\sqrt{13}}} & {{3 + \sqrt{13} \over 2}}\sqrt{4 \over {26 + 6\sqrt{13}}} 
		\end{pmatrix}}.
		\ee

		\noindent{We can absorb the positive
			numerical factors into $\Phi' ,\Psi'$ and the modified curvature scalar is given by:}
		
		\be
		{\mathcal{R}} = {R'}  - {1 \over 2}[{({{\nabla'}{\phi}})^2}
		- {({{\nabla'}{\psi}})^2}].
		\ee

		\noindent{The above expression gives a more convenient way to discuss the physical effects of the non-metricity scalar fields $\omega$ and $\psi$. We can express ${\tilde C}^{\alpha}_{~\mu \nu}$ in terms of $\phi$ and $\psi$ using the inverse of Eq.(58). We find that $\psi$ contributes a negative stress tensor to Einstein's equation and is a phantom scalar [6] of geometric origin. We always get a scalar field with a negative stress tensor in Einstein's equation, irrespective of the initial linear combinations of $\nabla'^{\mu} \Omega$ and $\nabla'^{\nu} \Psi$, provided they give finite ${\tilde C}^{\alpha~\mu}_{~\mu}$. $\psi$ vanishes when ${\tilde C}^{\alpha}_{~\mu \nu}$ is traceless in the lower indices. We can introduce various potential terms like ${{\mathcal V}_1}(\phi)$, ${{\mathcal V}_2}(\psi)$ and as per requirement. ${{\mathcal V}_1}(\phi)$ and ${{\mathcal V}_2}(\psi)$ can give us a theory more general than that given by Eq.(55) and new effects in the scalar-tensor theories. The contracted Bianchi identity will remain valid for this theory. Thus, transforming back to $(\Omega, \Psi)$ we will get a theory with $(\Omega, \Psi)$ that preserves the contracted Bianchi identity.}

		We can consider a theory containing $\phi$, $\psi$ and the cosmological constant $\Lambda$. This will lead to a possible theory of inflation together with dynamic and spatially varying dark energy. Einstein's equation will be generalized to:
		
		\be
		G'_{\mu \nu} + {\Lambda}{g_{\mu \nu}} - {\Lambda'_{\mu \nu} (\phi, \psi)} = 8{\pi}{P'_{\mu \nu}(\chi_a )} 
		\ee

		\noindent{where ${\Lambda'_{\mu \nu} (\phi, \psi)}$ includes ${P'_{\mu \nu}(\Phi)}, {P'_{\mu \nu}(\Psi)}$ and potential terms for $\phi$ and $\psi$. The left hand side gives a generalization of the $\Lambda$CDM theory linear in curvature and consistent with the Bianchi identities. This is along the line of introducing the cosmological constant in Einstein's equation. ${{P'}_{\mu \nu}(\chi_a)}$ is the stress tensor of ordinary matter and radiation including the right handed neutrinos expressed using the Levi-Civita connection. We can construct new non-trivial solutions of the vacuum Einstein's equation [28,59,60], \textit{i.e}, for ${P'_{\mu \nu}(\chi_a )} = 0$, with different choices of ${\Lambda_{\mu \nu} (\phi, \psi)}$. An important issue is the stability of the theory. $\psi$ provides a negative kinetic term, thus behaving like a phantom scalar [6,61,62,63,64]. Phantom field is a possible candidate to explain dark energy [5,6]. A phantom field, interacting with ordinary matter and radiation, can lead to instability both in  classical and quantum theory. Stable classical theories are discussed by a few authors by adding suitable potential terms for the phantom fields [63,64]. There remains the issue of unstable vacuum [65,66]. In the present theory, $\psi$ does not couple with ordinary matter and scattering between the two sectors causing instability of the vacuum is not possible [51]. Processes of the form: $Vacuum \rightarrow 2 \gamma + 2 \psi$, where $\gamma$ is a photon and $\psi$ gives a phantom quanta, need careful analysis. The coupling between $g_{\mu \nu}$ and $\psi$, given by ${({{\nabla'}{\psi}})^2}$, can be put in the form ${\psi}{\nabla'^2}{\psi}$, vanishes for the classical solutions, thus prohibiting the decay of vacuum to a detectable on-shell spectrum. Total derivatives in the action have no effect on the classical states. ${\psi}{\nabla'^2}{\psi}$ becomes a polynomial in $\psi$ with a ${{\mathcal V}(\psi)}$ for the on-shell spectrum in a nonperturbative theory, and there is no coupling with the graviton. Thus, classically stable self-interacting theory of $\psi$ can give a stable quantum theory. A theory with static or nearly static $\psi$ can be useful for discussing dark energy. A scalar field with negative energy density is also required to construct the \textit{steady state} model of the expanding universe [67-70].}

		$\phi$ contributes a positive stress tensor and gives a stable theory. Both ${Q_{\mu}}$ and ${{\bar{Q}}_{\mu \alpha \beta}}$ are finite for $\phi$. Thus, we can consider a quantum theory with only $\phi$ present as the non-metricity field. In [51,52] we have considered theories that can contain more than one $\phi$ fields. Such a theory can be useful to explain both inflation and dark energy [74]. We can introduce quadratic self-interaction term in $\phi$. We can couple $\phi$ with ordinary matter by adding suitable interaction terms. Being of geometric origin, the coupling between $g_{\mu \nu}$ and $\phi$ does not involve $G$. The same remains valid for $\psi$. $\phi$ and $\psi$ are quantum gravitational in origin, and it is expected that they can play important roles in the very early universe. $g_{\mu \nu}, \phi, \psi$ should be treated similarly in discussing quantum effects like vacuum fluctuations. Various free field configurations of $\phi$ and $\psi$, their vacuum energies, quantum fluctuations and classical configurations with possible interactions can be useful to explain inflation, dark energy and re-bouncing model of the universe [6,49,59].

		Equation (61) leads to a generalization of the $\Lambda$CDM theory. We do not obtain the Newtonian theory in the weak field and slow motion limit when $\Lambda \neq 0$, [28]. This implies that $\Lambda$ is small. This is similar to the non-metricities produced by the pair ($\Phi$, $\Psi$) or ($\phi$, $\psi$) discussed in the previous section. In this context, $\phi$ and $\psi$ have the advantage of being dynamical and hence can produce significant classical and quantum effects even in a free theory. This is along the line of perturbative quantum field theory and Casimir effect. Various non-trivial vacuum solutions of Einstein's equation, with and without $\Lambda$ [59,60], imply that affine connection need not to have ordinary matter and radiation as its sources even in the classical theory. The same can remain valid for $\phi$ and $\psi$. This agrees with the comments in the previous paragraph and the observed energy composition of our universe if these fields are useful to explain inflation and dark energy. It is unexpected that $10 \%$ ordinary matter can produce $70 \%$ dark energy. We find that inflation and dark energy possibly indicate that we have to generalize the classical structures of spacetime in the quantum domain. This is not completely unexpected. A complete theory of quantum mechanics can be partly spacetime independent algebraic theory as suggested by Einstein and apparent in various quantum entanglement experiments [71,72,73]. We will discuss applications of Eq.(61) in cosmology in a forthcoming article. ${P'_{\mu \nu}(\chi_a)}$ will be multiplied by appropriate factors when we transform to non-geometrized ordinary units. Quantum corrections to the effective actions of various theories mentioned above, like that given by Eq.(61), will contain higher order curvature invariants evaluated using the Levi-Civita connection together with quantum corrections coming for various matter fields and possibly for non-metricity fields $\phi$ and $\psi$. The last set will depend on the requirement of potential terms for $\phi$ and $\psi$. This is the case with some low energy effective actions of string theory that contain ghost field [6]. Lastly, a possible approach to explaining dark energy could be to consider $\psi$ as the possible source of $\Lambda$. Discussions on possible sources of $\Lambda$ for constant curvature spaces can be found in [59]. These are similar to $\psi$. We cannot generate the cosmological constant with perfect fluids for which both energy density and pressure are positive definite. We will discuss this aspect later. We can try to introduce additional fields using ${\tilde C}^{\alpha}_{~\mu \nu}$. This is discussed in [51,52].

\section*{VIII. Conclusion}

In this article, we have considered the issue of construction of covariant derivative operator in 
quantum gravity. We have used the canonical quantization approach and
Palatini action to illustrate this issue. 
We have found it is more perceptive to use all basic covariant structures of geometry 
to formulate a quantum theory of gravity.  
This is valid irrespective of the presence and nature of sources.
These covariant structures of geometry include metric
and a third rank tensor ${{C}^{\alpha}_{{~\mu \nu}}}$. We call this field as supertorsion. 
The later field leads to affine connections more general than
the metric compatible Levi-Civita connections. Symmetric part of ${C^{\alpha}_{~\mu \nu}}$ in the lower
indices can introduce scalar fields and symmetric second rank covariant tensors.
Antisymmetric part of this tensor in the lower indices 
gives half of torsion tensor.

We have found that the familiar Palatini formalism 
and metric-affine gravity are not sufficient to construct a quantum
theory of gravity. We have considered possible extensions of these formalisms to
construct a quantum theory. We can do so by using potentials to express connections. 
Alternatively, we can use more general actions than the Palatini action
to construct a quantum theory. We can include higher order curvature inavariants
like ${R_{\mu \nu}}{R^{\mu \nu}}$ in the metric-affine action. We can also use the Palatini 
$f(R)$ gravity or metric-affine $f(R)$ gravity. 
In the simple cases considered in this article, the potential formalism introduce two massless scalar fields
in the theory. They are non-localized and one of them contribute 
negative source stress-tensor to the familiar Einstein equation. 
This can be useful to explain early universe inflation and dark energy
that require negative pressure. This is also important
for wormholes and warp drive which require negative energy. 
The other scalar field, when supplemented by suitable potentials, 
will be useful to explain the early universe inflation [56,57].
We can introduce other fields from general affine connections.
We have found that general affine connections do not preserve the light cone under parallel transport 
and bring us beyond a strict local Minkowski spacetime. Inexactness of parallel transport 
in quantum gravity mentioned below Eq.(16) 
also does not preserve the light cone under parallel transport. Inflation, dark energy, dark matter,
renormalizability and the issue of $(3 + 1)$-splitting of spacetime 
into space and time have been primary fields of quantum gravity research that have found a common
place here. We find that the Lagrangian formalism, 
where we do not need $(3 + 1)$-splitting of spacetime into space and time,
is more general to construct a quantum theory of gravity.
We have also discussed the stability of the theory with a negative stress-tensor.
Affine connections can also be useful to construct theories
alternatives of cosmic inflation [59].

\section*{Acknowledgement}

I am thankful to a few reviewers for some improvements.

\vspace{0.5cm}

It is now great fun to see the half-leg shaped 'bridge', requiring a ladder to ride on it, at S.I.N.P (Kolkata). It is vividly colorful, large enough to be distinctly visible and was erected around 2001.

\section*{References}

[1] A. A. Starobinisky, Phys. Lett. B {\bf 91}, 99 (1980).

[2] A. H. Guth, Phys. Rev. D {\bf 23}, 347 (1981).

[3] D.N. Spergel \textit{et al} [WMAP Collaboration], Astrophys. J. Suppl. {\bf 148}, 175 (2003). 

[4] D.N. Spergel \textit{et al} [WMAP Collaboration], Astrophys. J. Suppl. {\bf 170}, 377 (2007).

[5] E. Komatsu \textit{et al} [WMAP Collaboration], Astrophys. J. Suppl. {\bf 180}, 330 (2009).

[6] L. Amendola and S. Tsujikawa, Dark Energy, (Cambridge University Press, Cambridge, 2010).

[7] F. Zwicky, Helv. Phys. Acta {\bf 6}, 110 (1933).
 
[8] A. De Felice and S. Tsujikawa, Living Reviews in Relativity. {\bf 13} (2010).

[9] D. N. Vollick, Phys. Rev. D {\bf 68}, 063510 (2003).

[10] S. Weinberg, The Cosmological Constant Problem, Rev. Mod. Phys. {\bf 61}, 1 (1989).

[11] Y. Fujii, Phys. Rev. D {\bf 26}, 2580 (1982).

[12] T. Chiba, T. Okabe and M. Yamaguchi, Phys. Rev. D {\bf 62}, 023511 (2000).

[13] A. Y. Kamenshchik, U. Moschella and V. Pasquier, Phys. Lett. B {\bf 511}, 265 (2001).

[14] S. Capozzillo, Int. J. Mod. Phys. D {\bf 11}, 483 (2002).

[15] L. Amendola, Phys. Rev. D {\bf 60}, 043501 (1999). 

[16] J. P. Uzan, Phys. Rev. D {\bf 59}, 123510 (1999).

[17] G. R. Dvali, G. Gabadadze and M. Porrati, Phys. Lett. B {\bf 485}, 208 (2000). 

[18] V. Sahni and Y. Shtanov, JCAP {\bf 311}, 014 (2003).

[19] R. Utiyama and B. S. DeWitt, J. Math. Phys. {\bf 3}, 608 (1962).  

[20] K. S. Stelle, Phys. Rev. D {\bf 16}, 953 (1977). 

[21] B. S. DeWitt, Phys. Reports {\bf 19}, No.6 (1975).

[22] N. D. Birrel and P. C. W. Davies, Quantum Fields in Curved Space 
(Cambridge University Press, Cambridge, 1982).

[23] I. L. Buchbinder, S. D. Odintsov, and I. L. Shapiro, 
Effective Actions in Quantum Gravity (IOP Publishing, Bristol, 1992). 

[24] G. A. Vilkovisky, Class. Quant. Grav. {\bf 9}, 895 (1992).

[25] J. G. Hocking and G. S. Young, Topology (Dover Publications, Inc., New York, 1961).

[26] D. Lovelock and H. Rund; Tensors, Differential Forms, and Variational Principals
(Dover Publications, Inc., New York, 1989).

[27] L. D. Landau and E. M. Lifshitz; The Classical Theory of Fields 
(Butterworth-Heinenann, Oxford, 1998).
 
[28] R. M. Wald, General Relativity (The University of Chicago Press, Chicago and London, 1984).

[29] C. W. Misner, K. S. Thorne and J. A. Wheeler, Gravitation (W.H. Freeman and company, New York, 1970). 

[30] S. W. Hawking: The Path-Integral Approach to Quantum Gravity \textsl{in} 
S. W. Hawking and W. Israel, eds, General Relativity: An Einstein Centenary Survey
(Cambridge University Press, 1979).

[31] A. Ashtekar, Lectures on Non-Perturbative Canonical Gravity (World Scientific, Singapore, 1991).

[32] F. W. Hehl \textit{et al}, Phys. Rept. {\bf 258}, 1 (1995).

[33] D. Iosiﬁdis, Class. Quant. Grav. {\bf 36}, 8 (2019).   

[34] M. Ferraris, M. Francavigilia and C. Reina, Gen. Rel. Grav. {\bf 14}, 243-254, (1982).

[35] V. Moncrief, J. Math. Phys. 16, 493.

[36] C. Isham: Canonical Quantum Gravity and the Question of Time, \textsl{in} J. Ehlers and 
H. Friedrich, eds, Canonical Gravity: From Classical to Quantum; 
Proceeding of the 117th WE Heraeus Seminar Held at Bad Honnef, Germany,1993 (Springer-Verlag, 1994).

[37] K. Sundermeyer, Constrained Dynamics (Springer-Verlag, 1995).
 
[38] F. W. Hehl \textit{et al}, Rev. Mod. Phys. {\bf 48} (1976) 3641.

[39] F. De Felice and C.J.S Clarke, Relativity on Curved Manifolds (Cambridge University
Press, Cambridge, 1990).

[40] C. Itzykson and J. B. Zuber, Quantum Field Theory (Dover Publications, Inc. Mineola, 2005) 

[41] P. Ramond, Field Theory: A Modern Primer (Addison Wesley, 1990).

[42] R. T. Hammond, Rep. Prog. Phys. {\bf 65}, 599 (2002).

[43] N. Koloper, Phys. Rev. D {\bf 44}, 2380 (1991).

[44] A. Golovnev, M. Mukhanov and V. Manchurin, JCAP {\bf 0806:009}, 2008.

[45] R. Emami \textit{et al}, JCAPO3 (2017) 058.

[46] N. Bartolo \textit{et al}, Phys. Rev. D {\bf 97}, 023503 (2018).

[47] D.E. Nevill, Phys.Rev. {\bf 23}D (1981) 1244; {\bf 25}D (1982) 573.

[48] E. Sezgin and P. van Nieuwenhuizen, Phys. Rev. {\bf 21}D (1981) 3269.

[49] J. V. Narlikar, Pramana {\bf 2}(3): 158–170 (1974).

[50] L. H. Ford and T. A. Roman, Scientific American {\bf 282} 46 (2000)

[51] K. Ghosh: Affine Connection, Quantum theory and New Fields,  Quantum Studies: Mathematics and Foundations, DOI :10.1007/s40509-024-00340-9.

[52] K. Ghosh: https://hal.science/hal-04086265.

[53] K. Ghosh: Affine connections in quantum gravity and new scalar fields. Physics of the Dark Universe, \textbf{26} (2019) 100403; https://hal.science/hal-02105422v4

[54] M. Gasperini and G. Veneziano, Phys. Rept. 373, 1 (2003).

[55] M. Nakahara, Geometry, Topology and Physics (Adam Hilger, Bristol and New York, 1990).

[56] S. Weinberg, The Quantum Theory of Fields, Vol.II (Cambridge University
Press, Cambridge, 1996).

[57] B. P. Abbott \textit{et al}, Phys. Rev. Lett. {\bf 119}, 161101, (2017).

[58] B. P. Abbott \textit{et al}, Astroph. J. {\bf 848}, L13, (2017). [1710.05834].

[59] S. W. Hawking and G. F. R. Ellis: The Large Scale Structure of Space-Time. Cambridge University Press, 1973.

[60] H. Stephani, et al.: Exact solutions of Einstein's field equations. Cambridge University Press, 2003.

[61] S. Dodelson and F. Schmidt, Modern Cosmology. Academic Press, 2021. 

[62] R. R. Caldwell, M. Kamionkowski, and N. N. Weinberg: Phantom Energy and Cosmic Doomsday, Phys. Rev. Lett. \textbf{91} (2003), 071301.

[63] S. M. Carroll, M. Hoffman and M. Trodden: Can the dark energy equation-of-state parameter $w$ be less than $-1$? Phys. Rev.D \textbf{70} (2003), 023509.

[64] P. Singh, M. Sami and N. Dadhich: Cosmological dynamics of phantom field. Phys. Rev.D \textbf{68} (2003), 023522.

[65] J. M. Cline, S. Jeon, and G.D. Moore, The phantom menaced: Constraints on low energy effective ghosts, Phys. Rev. D \textbf{70} (2004), 043543.

[66] R. R. Caldwell, M. Kamionkowski, and N. N. Weinberg: Phantom Energy and Cosmic Doomsday, Phys. Rev. Lett. \textbf{91} (2003), 071301.

[67] H. Bondi and T. Gold: The steady-state theory of the expanding universe. Mon. Not. Roy. Ast. Soc. \textbf{108} (1948), 252-70.

[68] F. Hoyle: A new model for the expanding universe. Mon. Not. Roy. Ast. Soc. \textbf{108} (1948), 372-82.

[69] F. A. E. Pirani: On the energy-momentum tensor and the creation of matter in relativistic cosmology, Proc. Roy. Soc. Lond. A \textbf{228} (1955), 455-62. 

[70] F. Hoyle and J.V. Narlikar: A new theory of gravitation. Proc. Roy. Soc. Lond. A \textbf{277} (1964), 1-23.

[71] J. Bub: Interpreting the Quantum World. Cambridge University Press, 1997.

[72] A. Fine: The Shaky Game. The University of Chicago Press, 1996.

[73] A. Aspect: Closing the door on Einstein and Bohr's quantum debate. Physics \textbf{8}, 123 (2015).

[74] N. J. Poplawski, Phys. Lett. B {\bf 694}, 181 (2010).

\section*{Appendix:A}

Here, we make a few comments on what will happen to Eq.(20) when we choose a different operator ordering
in the Levi-Civita connections than that discussed below Eq.(16). 
We first discuss what happens to the dicussions below
Eq.(20) when we consider a symmetric ordering in the Levi-Civita symbol.
The \textit{r.h.s} of Eq.(20) is now non-vanishing and we get the following 
condition when we consider the action of ${\hat{\nabla}'}_{x k}$ on both
sides of Eq.(19):

\be
-{\Big [}[\hat{g}^{\alpha \tau}(t,\vec{x}), \hat{M}_{\tau k (p}(t,\vec{x})]{\hat{g}_{|\alpha| l)}(t,\vec{x})},
\hat{\pi}^{p l}(t,\vec{y}) {\Big ]}
= {i}{{\hat{\nabla}'}_{x k}}[{\delta^{p}_{(p}}{\delta^{l}_{l)}}{{\delta}(\vec{x},\vec{y})}]
\ee

\noindent{Where, $ M_{\alpha \mu \nu} =   
	{1 \over 2}[{\partial_{\mu}}(g_{\alpha \nu}) + {\partial_{\nu}}(g_{\mu \alpha})
	- {\partial_{\alpha}}(g_{\mu \nu})]$ and we have kept 
	${\hat{g}_{\alpha l}}$ and ${\hat{g}_{\alpha p}}$ at the right
	of the connections. We have used the fact that ${\hat{\nabla}'}_{\mu}[\hat{g}_{\alpha \beta}] \equiv 0$,
	when the operator ordering in the Levi-Civita connections is taken to be the same as
	that discussed below Eq.(16) and given by Eq.(7). We also have, 
	${\hat{g}^{\alpha \kappa}}{\hat{g}_{\kappa \beta}} = {{\delta^{\alpha}_{\beta}}}$. 
	We need to solve the constraints and gauge fixing conditions to find exact 
	expressions of both sides. We can construct a general form as follows. 
	The \textit{r.h.s} was mentioned below Eq.(19).
	The commutator within the commutator in the \textit{l.h.s}
	of the above equation, when finite, will be independent of $\vec{y}$ and will 
	contain singular quantities in coordinates $\vec{x}$. 
	If we multiply both sides by a regular function $f(\vec{y})$ and integrate
	\textit{w.r.t} $\vec{y}$ over the spatial section, the \textit{r.h.s} will
	give a regular expression although the \textit{l.h.s} remains divergent.
	In the most favorable situation, the \textit{l.h.s} can be of the form:
	${i}s(\vec{x}, \vec{x}){{\hat{\nabla}'}_{x k}}[{\delta^{p}_{(p}}{\delta^{l}_{l)}}
	{\delta}(\vec{x},\vec{y})]$,
	where $s(\vec{x}, \vec{x})$ is a singular quantity. This is not same as the \textit{r.h.s}. 
	Thus, we again obtain a contradiction similar to that discussed below Eq.(20).  
	Lastly, many other ordering of $\partial_{\lambda}{g_{\alpha \beta}}$ 
	and $g^{\mu \nu}$ in the Levi-Civita connections can be expressed as a linear
	combination of the ordering given by Eq.(7) and the symmetric ordering
	considered here. Corresponding covariant derivative is given as:
	${\hat{\nabla}_{\mu}} = m {\hat{\nabla}_{1 \mu}} + (1 - m) {\hat{\nabla}_{2 \mu}}$, where 
	the covariant derivatives in the \textit{r.h.s} correspond to the two orderings mentioned before and
	$m$ can be negative. With ${\hat{\nabla}_{1 \mu}}({\hat{g}_{\alpha \beta}}) = 0$,
	we will again have contradictions similar to those discussed in this article.}

\end{document}